\documentclass[a4paper,11pt]{article}
\usepackage{jinstpub}
\usepackage{hyperref}
\usepackage{orcidlink}
\usepackage{comment}
\usepackage{caption}
\usepackage[toc,page]{appendix}
\usepackage{csquotes}
\usepackage{subcaption}
\usepackage{amsmath}
\usepackage{graphicx}
\usepackage{makecell}

\title{Response of wavelength-shifting and scintillating-wavelength-shifting fibers to ionizing radiation}

\author[a,1]{W.~Bae\orcidlink{0000-0002-7646-7577}\note{Corresponding author.},}
\author[a]{J.~Cesar\orcidlink{0000-0001-6644-0023},}
\author[a]{K.~Chen\orcidlink{0009-0008-5642-7624},}
\author[a]{J.~Cho\orcidlink{0009-0004-7530-7231},}
\author[a]{D.~Du\orcidlink{0009-0003-5668-2282},}
\author[a]{J.~Edgar\orcidlink{0009-0005-6003-8689},}
\author[a]{W.~Earthman\orcidlink{0009-0004-2040-953X},}
\author[b]{O.M.~Falana\orcidlink{0000-0003-0158-1997},}
\author[a]{M.~Gajda\orcidlink{0009-0002-2147-0848},}
\author[b]{C.~Hurlbut,}
\author[b]{M.~Jackson,}
\author[a]{K.~Lang\orcidlink{0000-0003-1269-7223},}
\author[a]{C.~Lee\orcidlink{0009-0003-5915-3642},}
\author[a]{J.Y.~Lee\orcidlink{0009-0008-4809-3775},}
\author[a]{E.~Liang\orcidlink{0009-0008-6995-3842},}
\author[a]{J.~Liu\orcidlink{0009-0009-4037-2179},}
\author[b]{C.~Maxwell,}
\author[a]{C.~Murthy\orcidlink{0000-0001-9044-7946},}
\author[a]{D.~Myers\orcidlink{0000-0001-8402-7240},}
\author[a]{S.~Nguyen\orcidlink{0009-0009-9622-6429},}
\author[a]{D.~Phan\orcidlink{0000-0002-0649-8167},}
\author[b]{T.~O’Brien,}
\author[a]{M.~Proga\orcidlink{0000-0002-0303-5159},}
\author[a]{S.~Syed\orcidlink{0009-0000-6163-5534},}
\author[a]{M.~Zalikha\orcidlink{0009-0002-7045-6022},}
\author[a]{J.~Zey\orcidlink{0009-0008-3005-6642}}

\affiliation[a]{Department of Physics, University of Texas at Austin, 
1 University Station, Austin, TX 78712-0264, USA}
\emailAdd{wonseokb@utexas.edu}

\affiliation[b]{Eljen Technology, 1300 W.\ Broadway, Sweetwater, TX 79556, USA}

\abstract{ 
We report results of characterizing the response and light transport of wavelength-shifting (WLS) and scintillating-wavelength-shifting (Sci-WLS) fibers under irradiation by radioactive $\alpha$, $\beta$, and $\gamma$ sources. 
Light yield and light transmission were measured for the WLS fiber BCF-91A from Saint-Gobain and for a new Sci-WLS fiber EJ-160 from Eljen Technology. 
 
The two variants with different fluor mixtures, EJ-160I and EJ-160II, exhibited approximately five and seven times higher light yield than BCF-91A, respectively, while their attenuation lengths were 3.80\,m for BCF-91A, 4.00\,m for EJ-160I, and 2.50\,m for EJ-160II.
}

\keywords{ionizing radiation; light-guide fibers; wavelength-shifting; scintillating-wavelength-shifting fibers; photoelectron; light yield; SiPM}

\begin{document}
\maketitle
\flushbottom

\section{Introduction}

Plastic wavelength-shifting and scintillating fibers offer efficient light collection and transport; thus, they have found a broad range of applications in particle and nuclear physics experiments. Recent examples include MINOS~\cite{MINOS-Michael:2008bc, Avvakumov:2005ww}, NOvA~\cite{Ayres:2004js}, T2K~\cite{T2K-ND280-NIMA}, the LHCb tracker~\cite{LHCb-SciFi-NIMA2025}, and the GERDA and LEGEND-200 liquid argon veto system~\cite{Ackermann:2012xja, LEGEND:2025insdet}. In the future, the LEGEND-1000~\cite{LEGEND:2021bnm}, the ePIC~\cite{Klest:2024-CalorimetryEPIC} and other experiments plan to continue using and refining such techniques.

In some detector designs, fibers are not only required to provide efficient light collection and wavelength-shifting, but also to meet additional stringent requirements on radiopurity~\cite{Ackermann:2012xja, LEGEND:2025insdet, LEGEND:2021bnm}.
An attractive option is the use of fibers as radiation detectors of natural nuclear and cosmogenic radioactivity. For example, fibers immersed in liquid argon can detect $\beta$ decays of \textsuperscript{39}Ar or $\beta$ decays of \textsuperscript{42}K from cosmogenic \textsuperscript{42}Ar~\cite{Ackermann:2012xja, LEGEND:2025insdet}. 
Additionally, properly doped and formulated fibers can self-tag their own radio-impurities. This may enhance the veto power while relaxing the radiopurity requirement of the manufacturing process.

We have partnered with Eljen Technology~\cite{Eljen2} to develop new fibers that would not only be competitive with the fibers available on the market, but would be better optimized for the needs of upcoming particle physics experiments such as LEGEND-1000~\cite{LEGEND:2021bnm}.
%
In this work, we report the results of a comprehensive study of two new scintillating-wavelength-shifting (Sci-WLS) fibers from Eljen Technology named EJ-160I and EJ-160II and compare their performance with the wavelength-shifting (WLS) fiber  BCF-91A from Saint-Gobain~\cite{Saint-Gobain2}, now Luxium Solutions~\cite{Luxium-Solutions}. 

\section{Fiber samples}

\begin{figure}[h!]
    \setlength{\abovecaptionskip}{5pt}   
    \setlength{\belowcaptionskip}{0pt}   
    \centering
    \includegraphics[width=.495\textwidth, height=.37\textwidth]{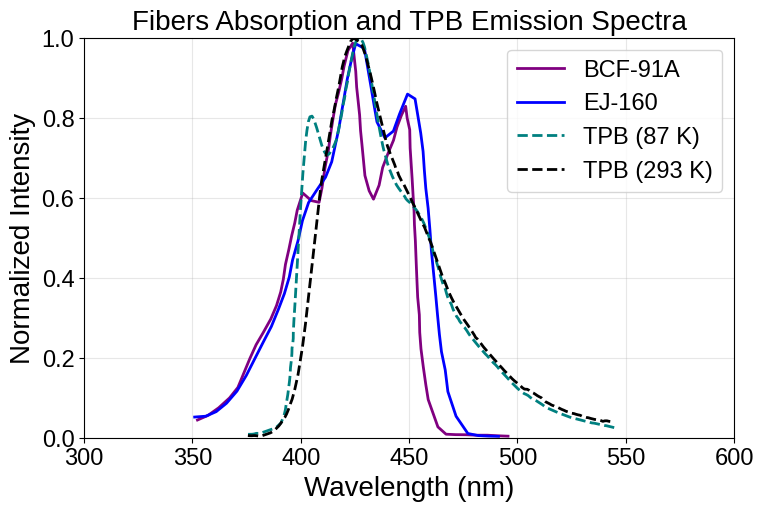}
    \includegraphics[width=.495\textwidth, height=.37\textwidth]{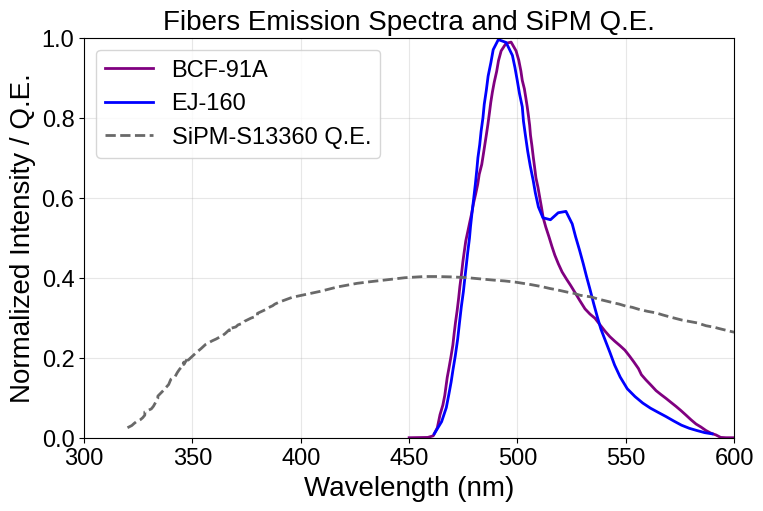}
    \caption{Absorption (left) and emission (right) spectra of the WLS fibers (manufacturers' data). EJ-160 has an absorbance peak at 427\,nm and an emission peak at 490\,nm, while BCF-91A has an absorbance peak at 424\,nm and an emission peak at 494\,nm.
    For reference, we include the emission spectra of tetraphenylbutadiene (TPB) from~\cite{TPB-Leonhardt-JINST-2024}, which is often used for shifting scintillation light of liquid argon or liquid xenon, and the quantum efficiency of Silicon Photomultiplier (SiPM) S13360 from Hamamatsu Photonics~\cite{hamamatsu} 
    All fibers and TPB spectra are normalized to their respective maxima.
    Figure adapted from~\cite{Bae:2025fiberLED}.
    }
    \label{fig: fiber emission spectrum1}
\end{figure}

\begin{figure}[h!]
\centering
\includegraphics[width=.95\textwidth]{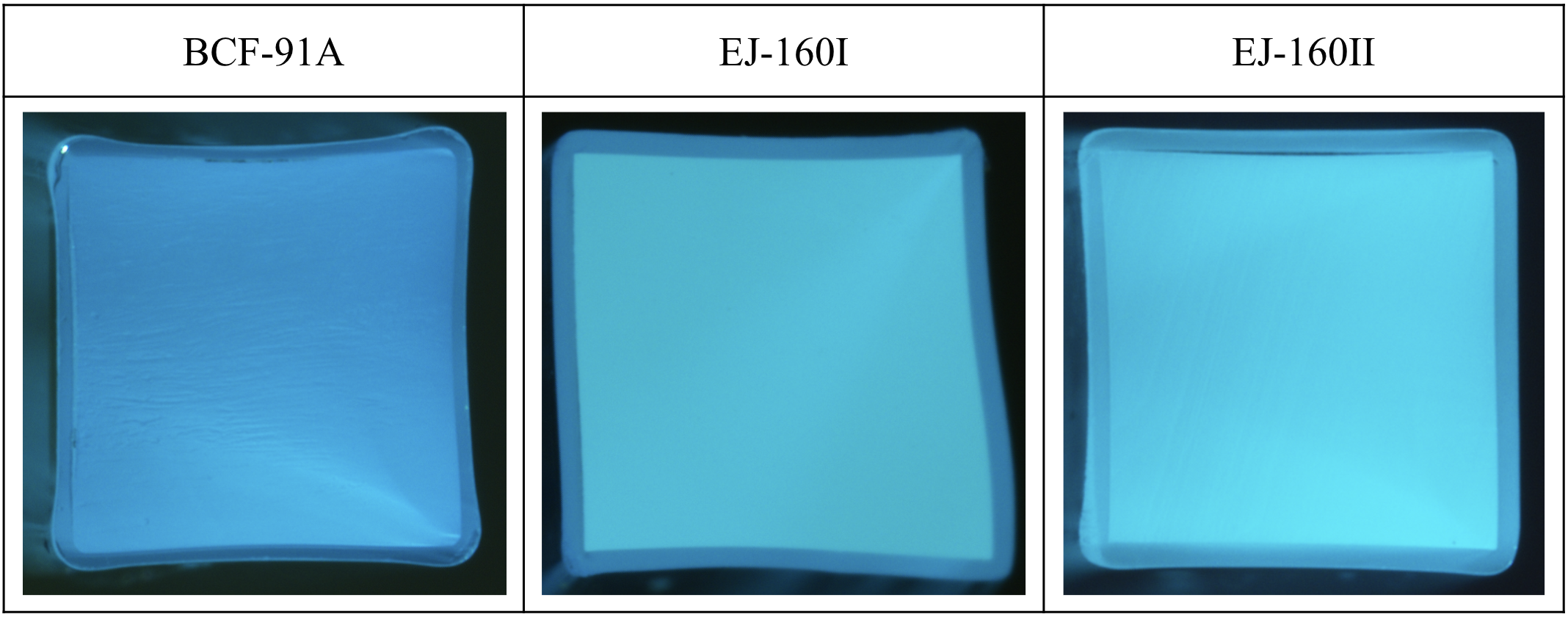}
\caption{\raggedright
Pictures of diamond fly-cut cross sections of the three tested fibers. These images were captured under a microscope with external illumination to highlight the core/cladding boundaries. Figure adapted from~\cite{Bae:2025fiberLED}.
}
\label{fig: microscope inspection}
\end{figure}

Table \ref{table:wls_fibers1} and Figure~\ref{fig: fiber emission spectrum1} summarize the main physical and optical properties of three fibers: BCF-91A from Saint-Gobain and the newly developed EJ-160 fibers from Eljen Technology, which are produced in two variants—EJ-160I and EJ-160II—featuring different fluor mixtures. The BCF-91A fiber type was previously used in GERDA~\cite{Ackermann:2012xja} and LEGEND-200 \cite{LEGEND:2025insdet}. 
The BCF-91A test samples were supplied to us by a group at the Technical University of Munich.
Figure~\ref{fig: microscope inspection} shows images of diamond fly-cut cross sections of fibers captured with a 
microscope~\cite{Nikon-microscpoe}. The BCF-91A sample has a single cladding of approximately 0.03\,mm thickness, while Eljen Technology fibers each have a cladding layer of 0.04\,mm thickness.

\begin{table}[h!]
\small
\centering
\begin{tabular}{|p{3.5cm}|c|c|}
    \hline
    \multicolumn{1}{|c|}{\textbf{Feature}} 
        & \textbf{BCF-91A} 
        & \textbf{EJ-160} \\
    \hline
    Variant 
        & standard 
        & \makecell{EJ-160I\\EJ-160II} \\
    \hline
    Type 
        & wavelength-shifting 
        & scintillating-wavelength-shifting\\
    \hline
    Cross-section 
        & 1~mm square 
        & 1~mm square \\
    \hline
    Cladding 
        & single\textsuperscript{(*)}  
        & single\textsuperscript{(*)} \\
    \hline
    Core material  
        & polystyrene 
        & polystyrene \\
    \hline
    Cladding material 
        & PMMA 
        & PMMA \\
    \hline
    \makecell[{{p{3.5cm}}}]{Refractive index \\ (core / cladding)}
        & 1.60/1.49 
        & 1.59/1.49 \\
    \hline
    Cladding thickness 
        & 0.03\,mm\textsuperscript{(*)} 
        & 0.04\,mm\textsuperscript{(*)} \\
    \hline
\end{tabular}

\caption{Properties of tested fibers. PMMA is polymethyl methacrylate. 
\textsuperscript{(*)} The thicknesses of claddings were measured by our microscope.}
\label{table:wls_fibers1}
\end{table}


\section{Experimental setup}
 
\begin{figure}[h!]
\centering
\includegraphics[width=.45\textwidth, height=.425\textwidth]{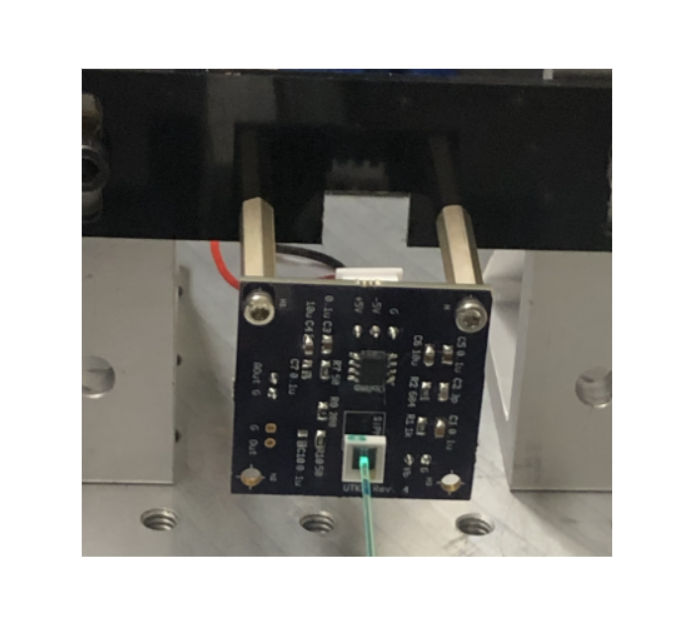}
\includegraphics[width=.45\textwidth, height=.40\textwidth]{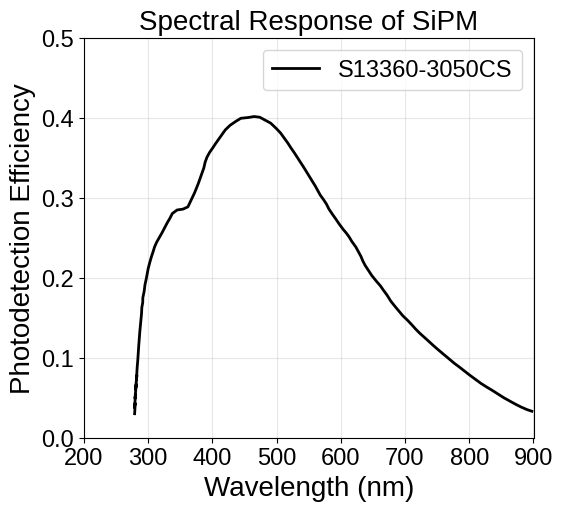}
\caption{\raggedright 
Left: SiPM readout board with the Hamamatsu Photonics SiPM S13360-3050CS coupled to a fiber. Fiber ends were polished using a diamond fly-cutter and coupled to the SiPM using optical grease. Right: Photon detection efficiency of the same SiPM~\cite{hamamatsu}.
}

\label{fig: Hamamatsu SiPM}
\end{figure}

\begin{figure}[h!]
\centering
\includegraphics[width=.45\textwidth, height=.4\textwidth]{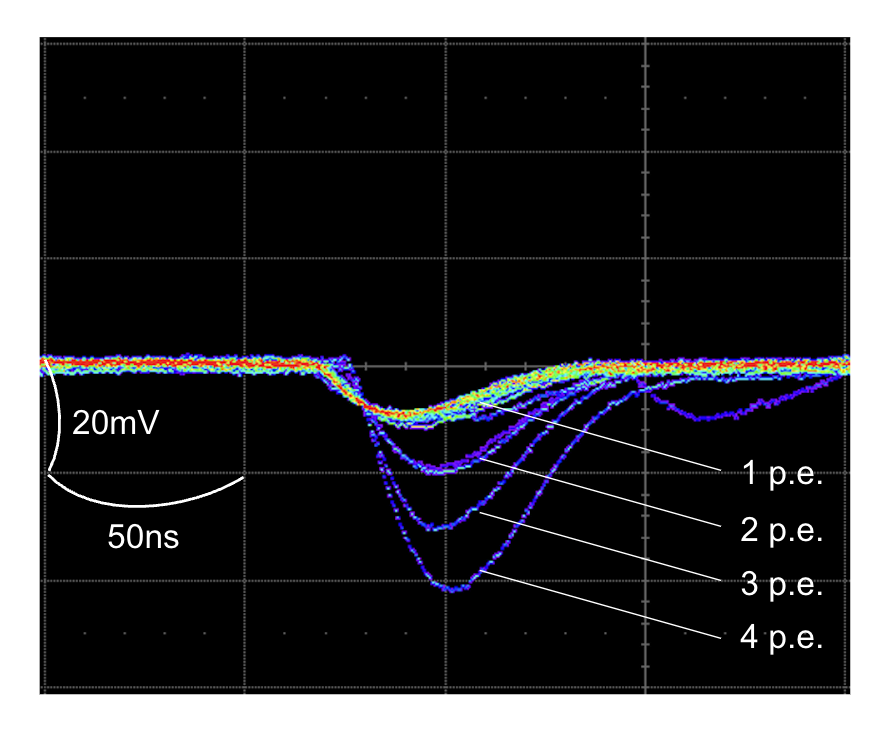}
\includegraphics[width=.54\textwidth, height=.4\textwidth]{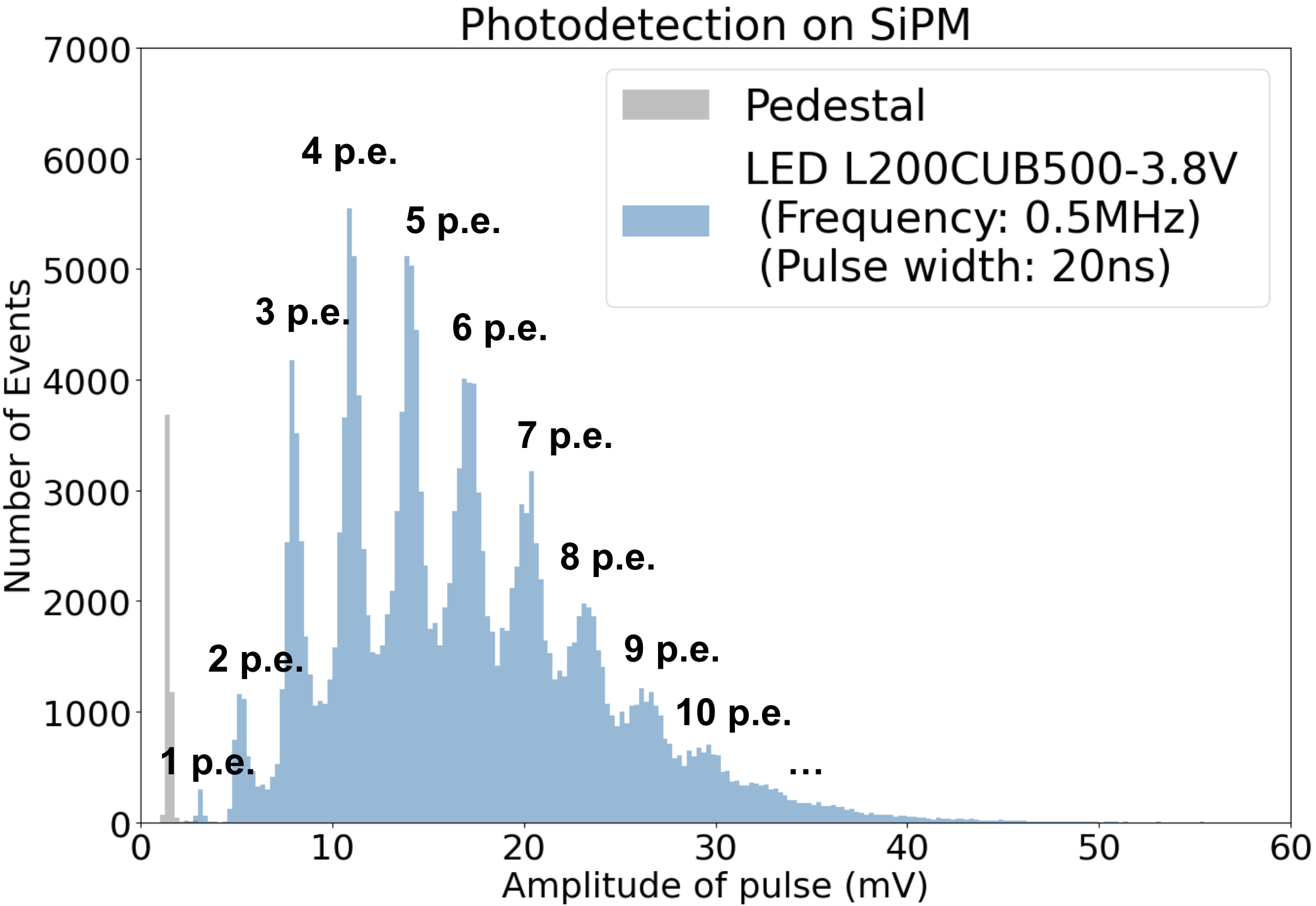}
\caption{\raggedright
Left: Typical SiPM signal traces recorded on the oscilloscope with a 0.5 photoelectron (p.e.) trigger threshold.%
\ Right: Histogram of SiPM pulse amplitudes (mV) obtained under LED excitation, yielding higher p.e.\ count. The legend shows the frequency of pulse generator and pulse width applied to the LED L200CUB500-3.8V from LEDtronics, Inc~\cite{LEDtronics}.}
\label{fig: sample pulse}
\end{figure}

The fibers were approximately 1.4\,m long. They all had both ends optically coupled to Hamamatsu Photonics S13360-3050CS SiPMs with a 3$\times$3~mm$^2$ active area and 50~$\mu$m microcells using optical grease BC-630 from Saint-Gobain~\cite{Saint-Gobain2}. 
Figure~\ref{fig: Hamamatsu SiPM} shows the SiPM mounted on the custom-designed readout board and presents its photon detection efficiency.
The readout board distributes the SiPM bias voltage, and the amplified output was provided by an on-board transimpedance stage with a 604~$\Omega$ feedback resistor and a 3~pF feedback capacitor.
The amplified output was digitized with a WaveRunner HRO 66Zi oscilloscope from Teledyne LeCroy~\cite{TeledyneLeCroy}.
Representative traces and the corresponding digitized pulse amplitude histogram are shown in Figure~\ref{fig: sample pulse}.

We used standard $\alpha$, $\beta$, and $\gamma$ radioactive calibration sources to irradiate fibers at different distances from the end. SiPM pulse-heights were recorded on the scope and stored data were analyzed to evaluate the light-yield and light transmission of the fibers. 

\section{Fiber response measurements}

\subsection{Beta response}
\label{sec:Beta_irradiation}

\begin{figure}[h!]
\centering
\includegraphics[width=.99\textwidth]{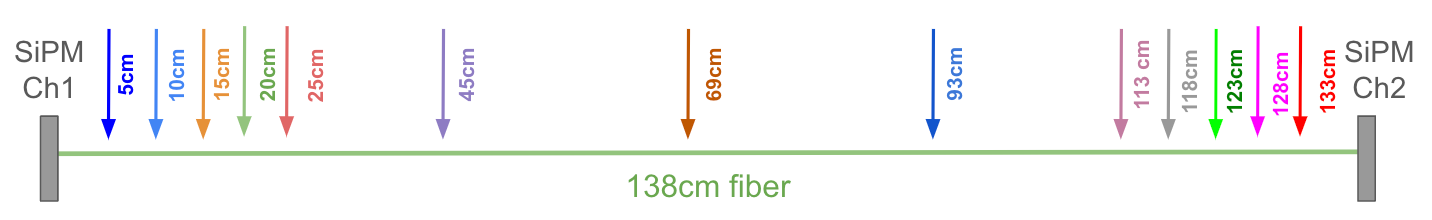}
\caption{\raggedright A schematic view of the setup for irradiation along a 138\,cm-long fiber coupled to SiPM's at both ends for $\beta$ and $\gamma$ irradiation studies. 
}
\label{fig: Optical_Setup1}
\end{figure}

For these measurements, a standard 1-inch-diameter plastic disk with embedded \textsuperscript{90}Sr isotope of activity \textcolor{black}{4.8\,$\mu$Ci} served as the $\beta$ (electron) radiation point source. 
The disk was mounted on a 0.1-inch thick steel collimator with a diameter of 0.1 inch, installed on an optical bench in a dark box.
This setup enabled nearly perpendicular irradiation of the fibers at 13 positions between 5\,cm and 133\,cm from one end, as schematically illustrated in Figure~\ref{fig: Optical_Setup1}.

\begin{figure}[h!]
\centering
\captionsetup[figure]{skip=5pt}

\begin{subfigure}{0.495\textwidth}
  \includegraphics[width=\linewidth,height=0.65\textwidth]{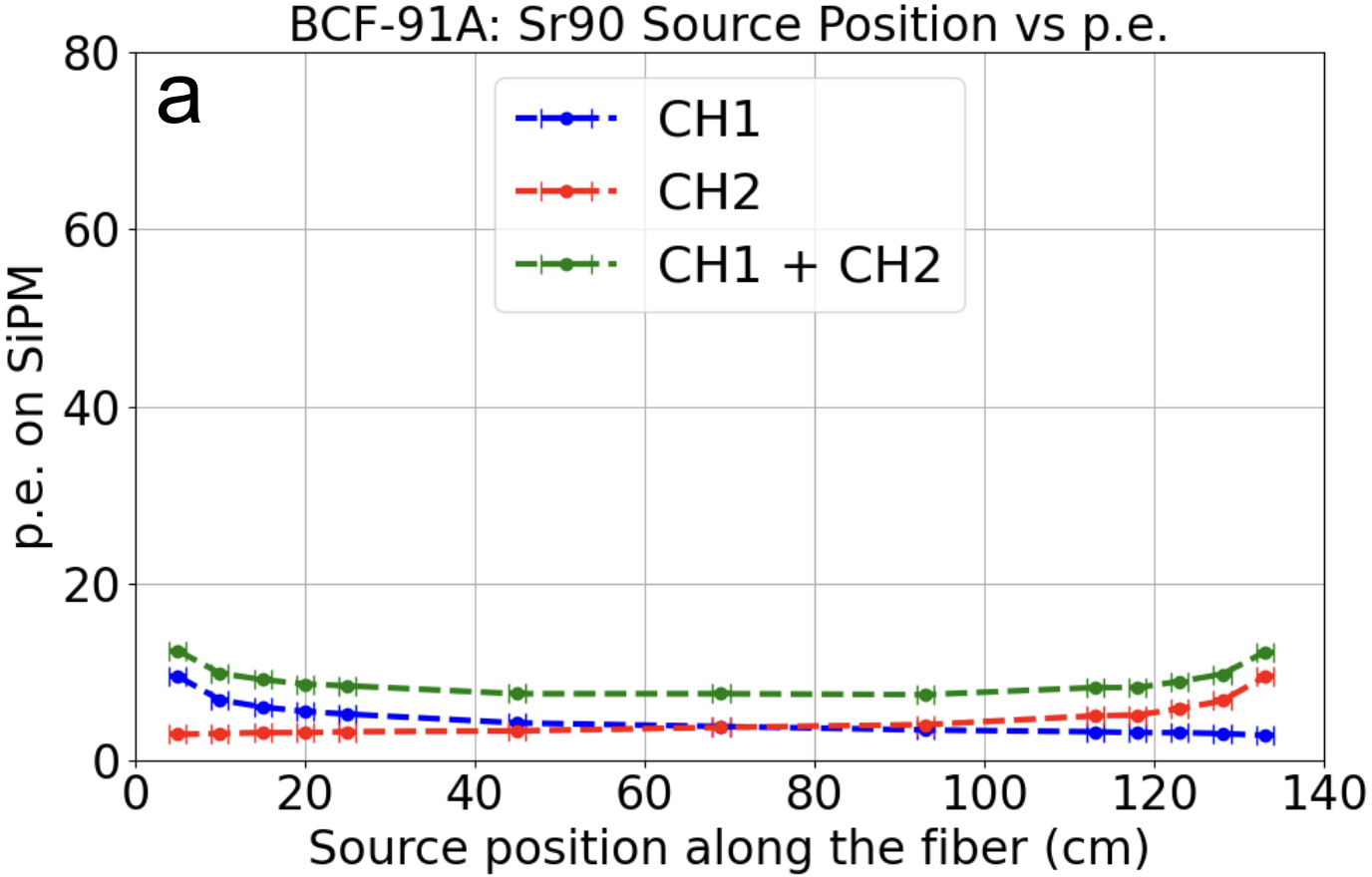}
\end{subfigure}
\hfill
\begin{subfigure}{0.495\textwidth}
  \includegraphics[width=\linewidth,height=0.65\textwidth]{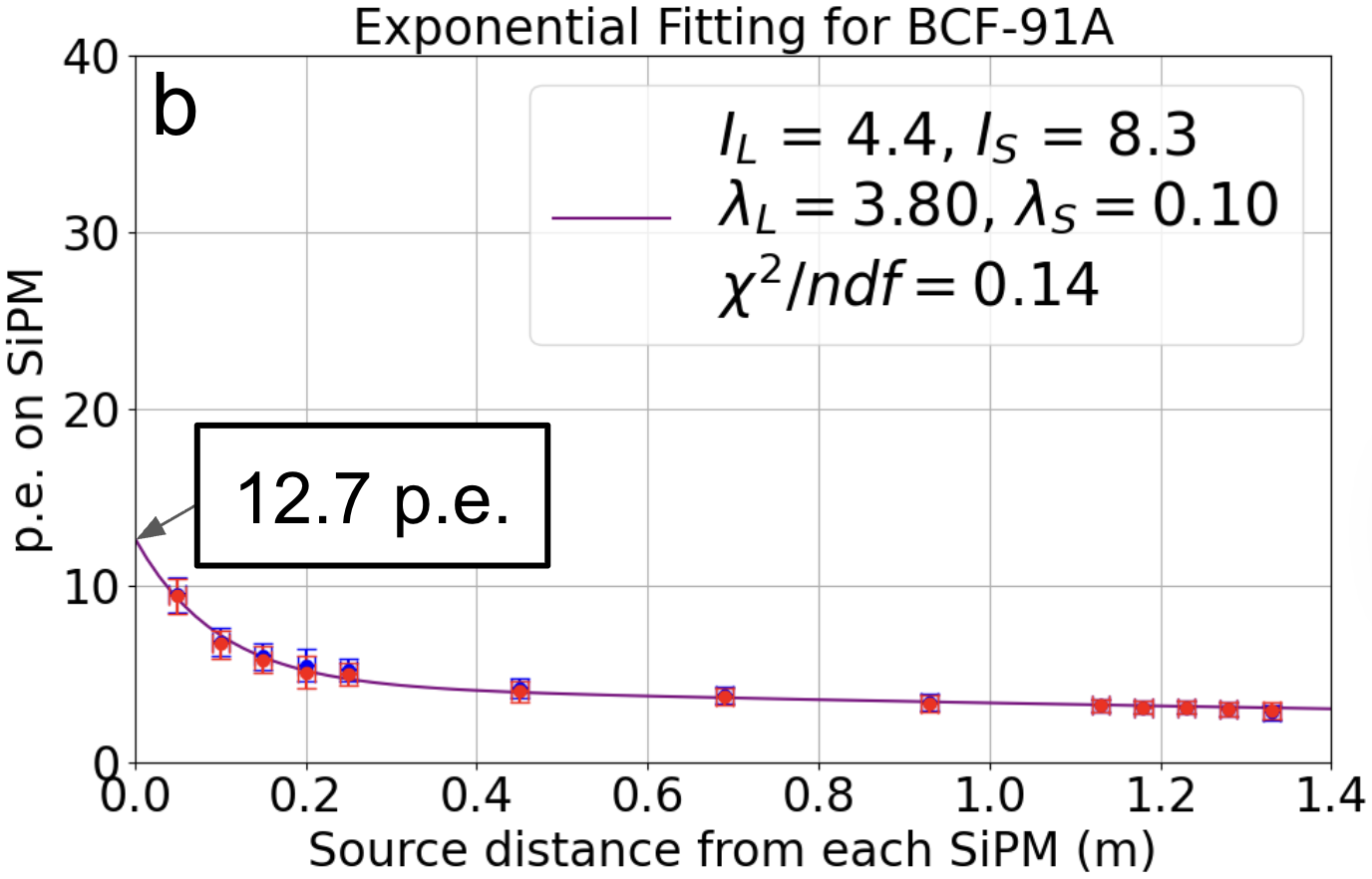}
\end{subfigure}

\vspace{1ex}

\begin{subfigure}{0.495\textwidth}
  \includegraphics[width=\linewidth,height=0.65\textwidth]{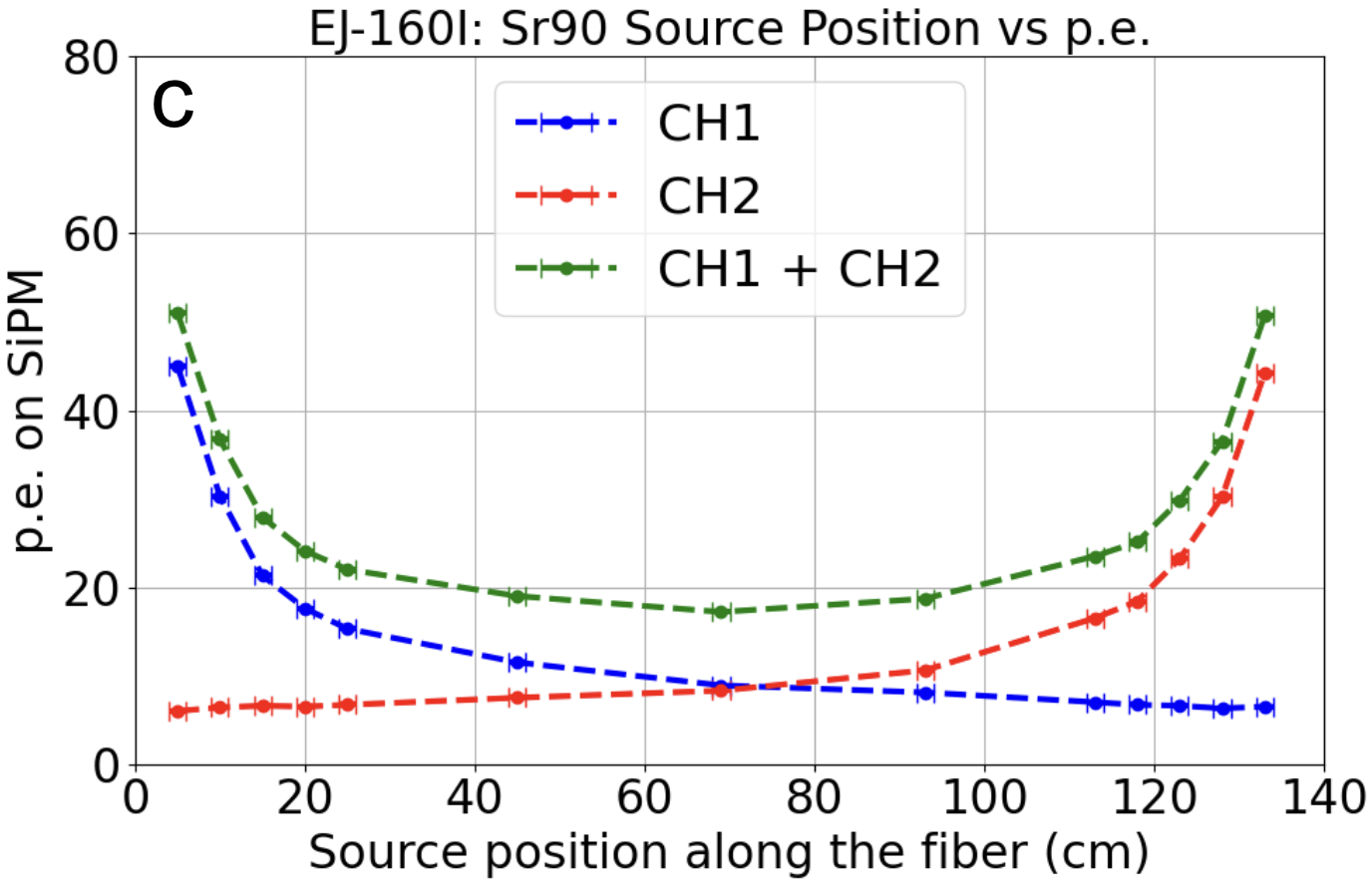}
\end{subfigure}
\hfill
\begin{subfigure}{0.495\textwidth}
  \includegraphics[width=\linewidth,height=0.65\textwidth]{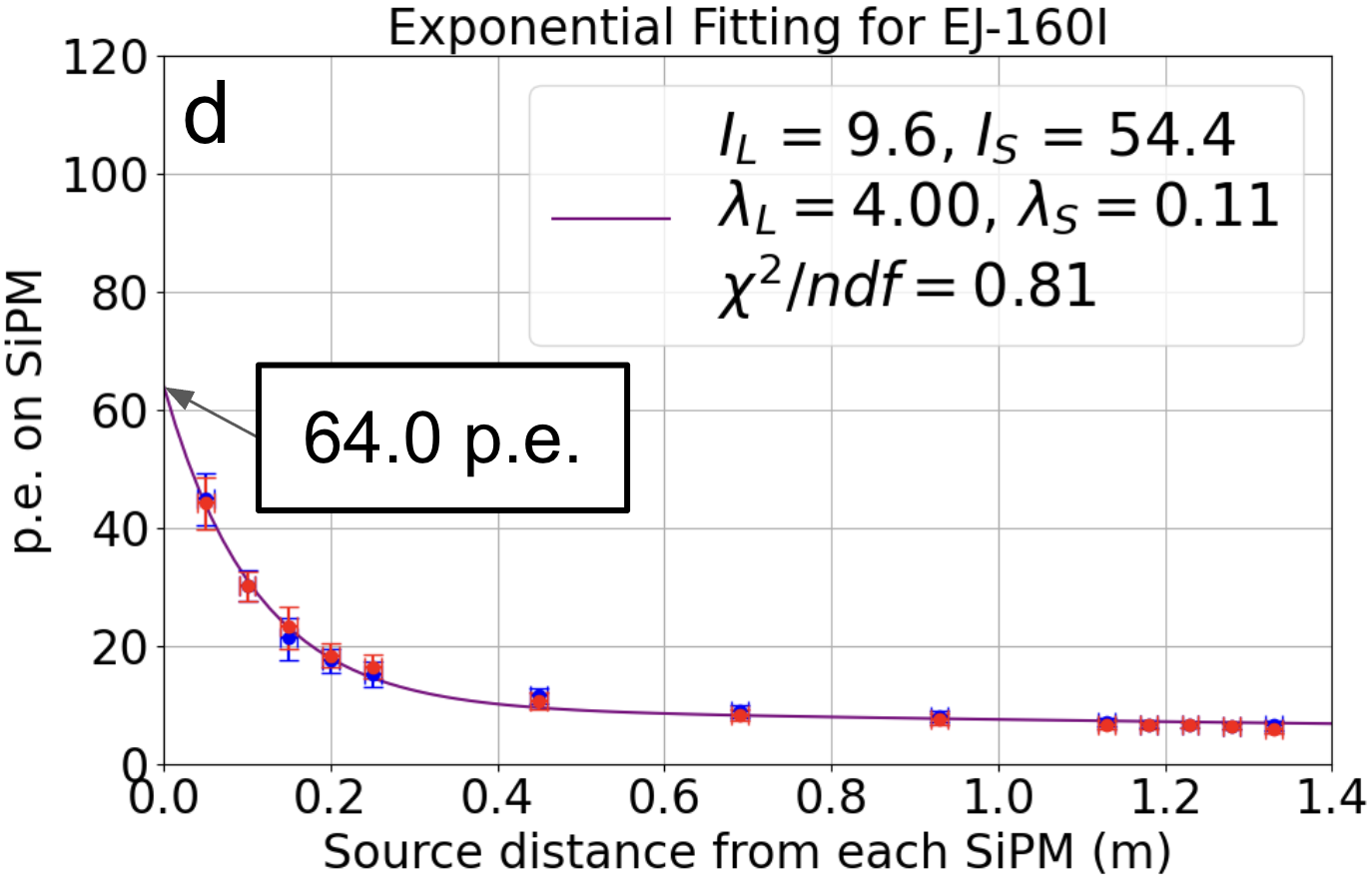}
\end{subfigure}

\vspace{1ex}

\begin{subfigure}{0.495\textwidth}
  \includegraphics[width=\linewidth,height=0.65\textwidth]{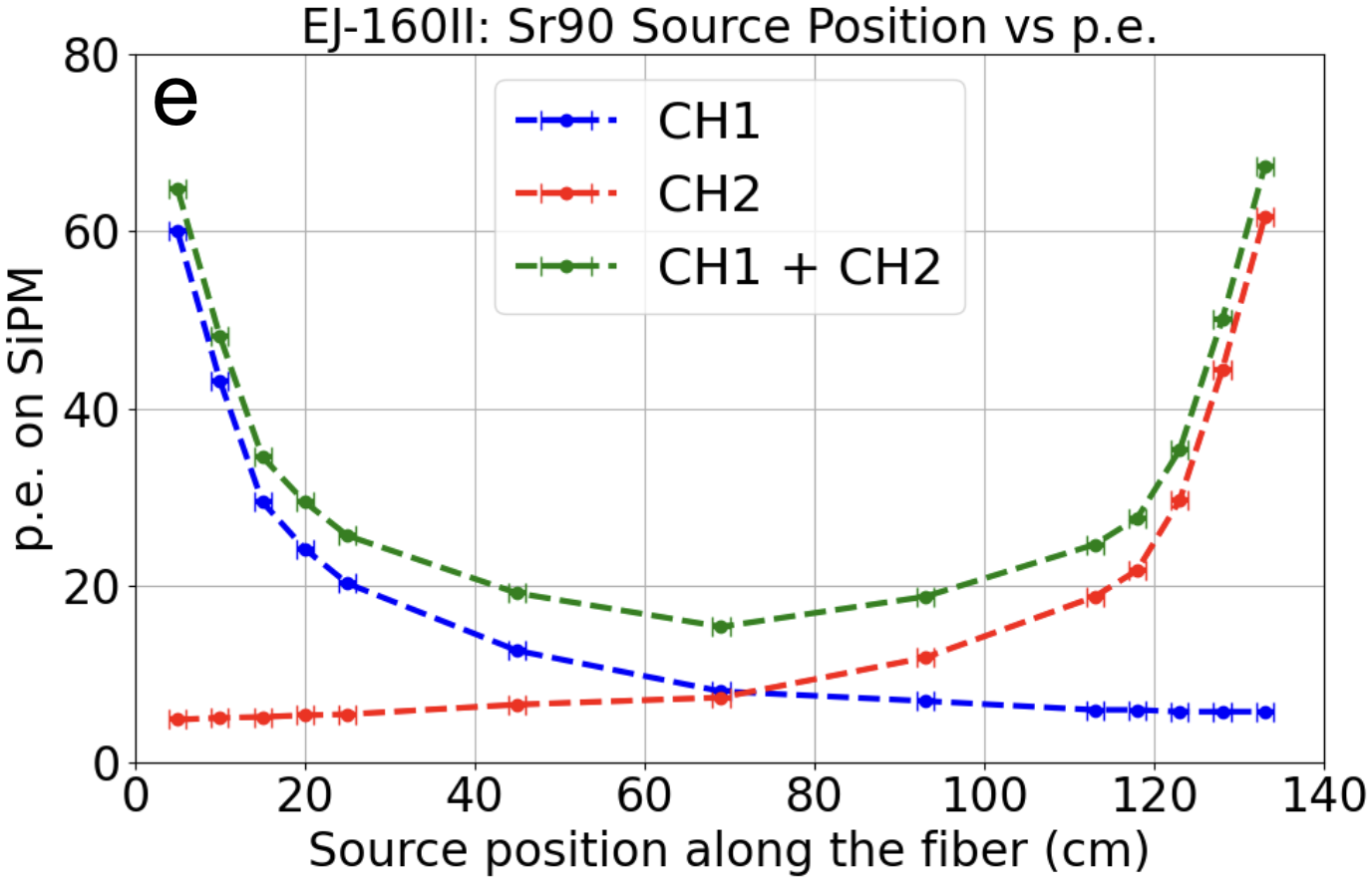}
\end{subfigure}
\hfill
\begin{subfigure}{0.495\textwidth}
  \includegraphics[width=\linewidth,height=0.65\textwidth]{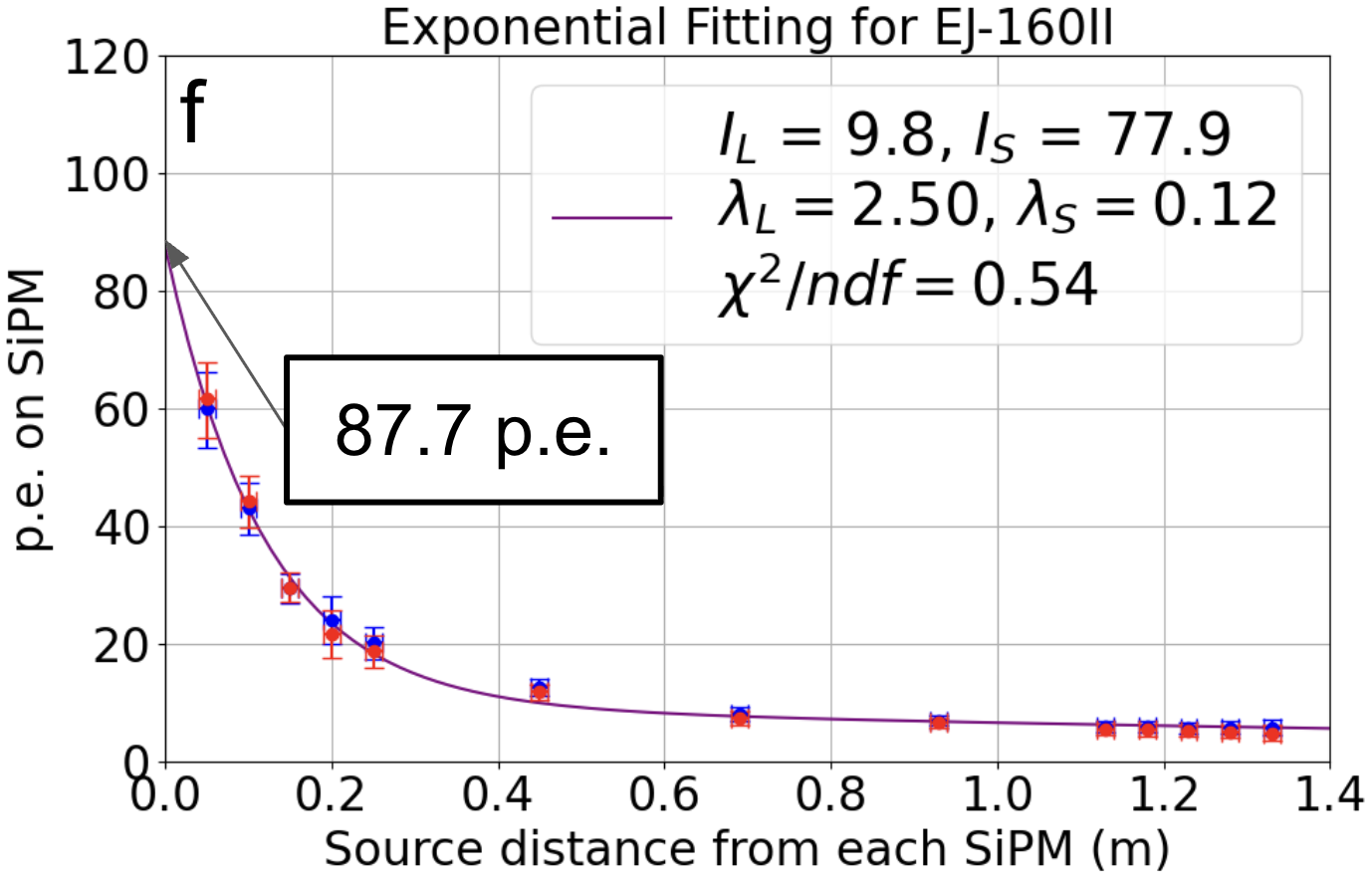}
\end{subfigure}

\caption{Light yields and fitted attenuation lengths of fibers irradiated with a \textsuperscript{90}Sr $\beta$ source. 
In the left panels, the blue, red, and green curves represent channel 1 (CH1; left SiPM), channel 2 (CH2; right SiPM), and their sum, respectively; the right panels show CH1 (blue) and CH2 (red) with the corresponding double-exponential fits.}

\label{fig: beta_sum_fitting}
\end{figure}

Figure~\ref{fig: beta_sum_fitting} presents the mean number of photoelectrons (p.e.) detected by SiPMs for the three fibers. 
A low statistical uncertainty of the mean is achieved by a 10-minute measurement at each source position, thus comprising tens of thousands of events.
The left column of the figure, panels (a), (c) and (e), shows raw mean data, 
while the right column, panels (b), (d) and (f), shows the corresponding double-exponential fits.
The fits follow the functional form:
\begin{equation}
I = I_{\text{long}}\,e^{-x/\lambda_{\text{long}}} + I_{\text{short}}\,e^{-x/\lambda_{\text{short}}},
\label{formula}
\end{equation}
where $I$ is the number of photoelectrons (p.e.) recorded by the SiPM, and $x$ is the distance between the radiation source and the SiPM. $I_{\text{long}}$, $I_{\text{short}}$, $\lambda_{\text{long}}$, and $\lambda_{\text{short}}$ denote the long and short components of the light yield and attenuation length.
An equivalent parameterization is also commonly written as
$I(x)=I_{0}\left[\alpha e^{-x/\lambda_{\text{long}}}+(1-\alpha)e^{-x/_{\text{short}}}\right]$,
where $\alpha$ is a constant mixing factor~\cite{Kodama2024}.
In this notation, $I_{0}\alpha$ and $I_{0}(1-\alpha)$ correspond to
$I_{\text{long}}$ and $I_{\text{short}}$, respectively, in Eq.~(\ref{formula}).
We have evaluated this approach, and it provided the same quality fit of our data.
In some cases, an effective single attenuation length is reported from a single-exponential fit over a selected distance range, often focusing on distances beyond the short-distance region (e.g., 0.3--1.2\,m~\cite{Alekseev2022}, or 1.0--2.8\,m~\cite{Barbosa2013}).
However, at shorter distances, a double-exponential function provides a better description of the light yield~\cite{MINOS-Michael:2008bc,Avvakumov:2005ww}.


For BCF-91A (Figure~\ref{fig: beta_sum_fitting}a), the blue, red, and green lines represent channel 1 (CH1) or SiPM left, channel 2 (CH2) or SiPM right, and their sum, respectively, exhibiting a symmetric dependence on the distance to the radioactive source. 
The light attenuation is well described by a double-exponential function as shown in Figure~\ref{fig: beta_sum_fitting}b.
  
By extrapolating the fits to the zero point of the x-axis (the source position relative to each SiPM), we estimate the light yield at $x = 0$ for a given fiber, following a framework similar to that established in previous studies~\cite{Moszynski2014, Moszynski2013}.

They range from 12.7\,p.e. for BCF-91A, 64.0\,p.e. for EJ-160I, and 87.7\,p.e. for EJ-160II.
The long attenuation lengths that are fitted using formula \ref{formula} are 3.80\,m (BCF-91A), 4.00\,m (EJ-160I), and 2.50\,m (EJ-160II). 

\begin{figure}[h!]  
    \centering  
    \includegraphics[width=0.75\textwidth]
    {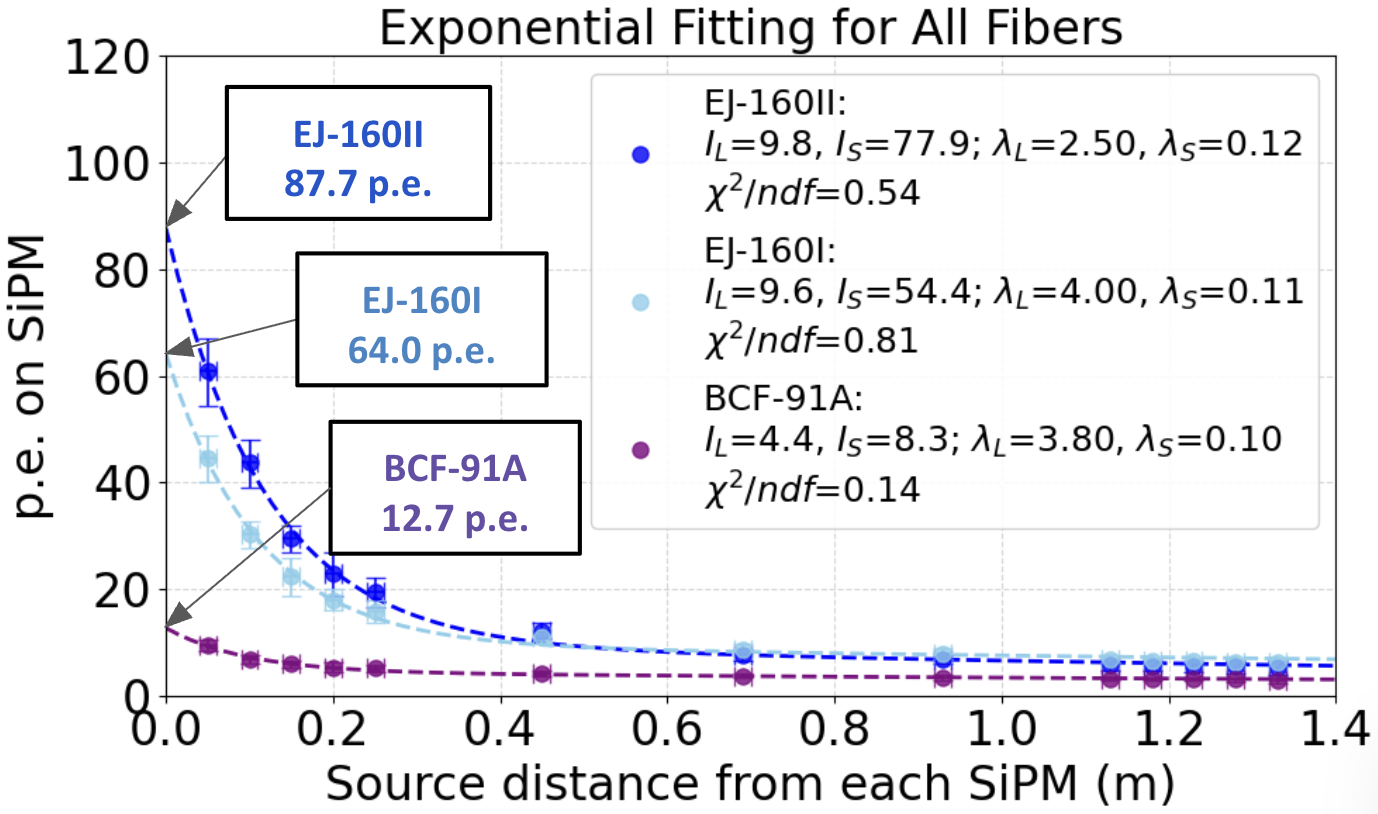}  
        \caption{
        Summary of the $\beta$ response for the three fibers tested.
        }
    \label{fig: beta_summary}  
\end{figure}

EJ-160II shows higher light yield at the cost of shorter attenuation lengths compared to the EJ-160I. 
In Figure~\ref{fig: beta_summary} we directly compare the response of the three fibers to $\beta$ irradiation.

The systematic uncertainties of these measurements were initially evaluated by repeating the measurements ten times ``from scratch'' (i.e., completely deconstructing and rebuilding the experimental setup) to project onto the overall systematic uncertainty of this technique.
This included resetting the fiber on the SiPMs by removing residual optical grease, inspecting the SiPM window for dust/debris, and reapplying optical grease before re-coupling, as well as repositioning the sources on the fibers under test.

%
The statistical uncertainty is effectively negligible since each measurement contains tens of thousands of events.

\subsection{Gamma response}\label{sec:gamma_irradiation}

\begin{figure}[h!]
\centering
\includegraphics[width=.99\textwidth]{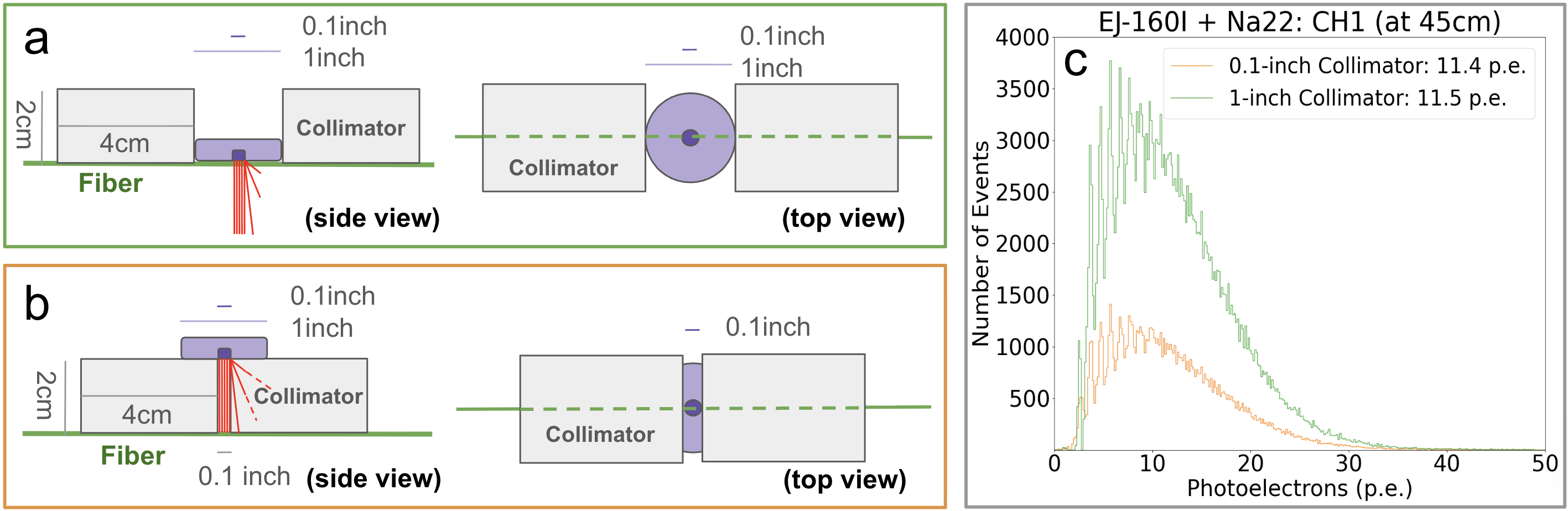}
\caption{\raggedright 
(a) A schematic view of the setup for $\gamma$ testing.
(b) A similar setup with narrower collimation. This setup was deemed impractical to collect large statistics.
(c) Green and orange curves show measurement using a collimation scheme shown in a and b, respectively.
}
\label{fig: Collimator - Gamma}
\end{figure}

For these measurements, a \textsuperscript{22}Na point source was used as a {$\gamma$} source of activity 3.4\,$\mu$Ci with the predominant 511\,keV gamma line. The same setup as for the $\beta$ studies was used, but the source was additionally collimated and shielded with copper, as illustrated in a schematic drawing in Figure~\ref{fig: Collimator - Gamma}a. A narrower collimation was attempted, as shown in Figure~\ref{fig: Collimator - Gamma}b, but was considered impractical to collect high-statistics data. 

\begin{figure}[h!]
\centering
\captionsetup[figure]{skip=5pt}

\begin{subfigure}{0.495\textwidth}
  \includegraphics[width=\linewidth,height=0.65\textwidth]{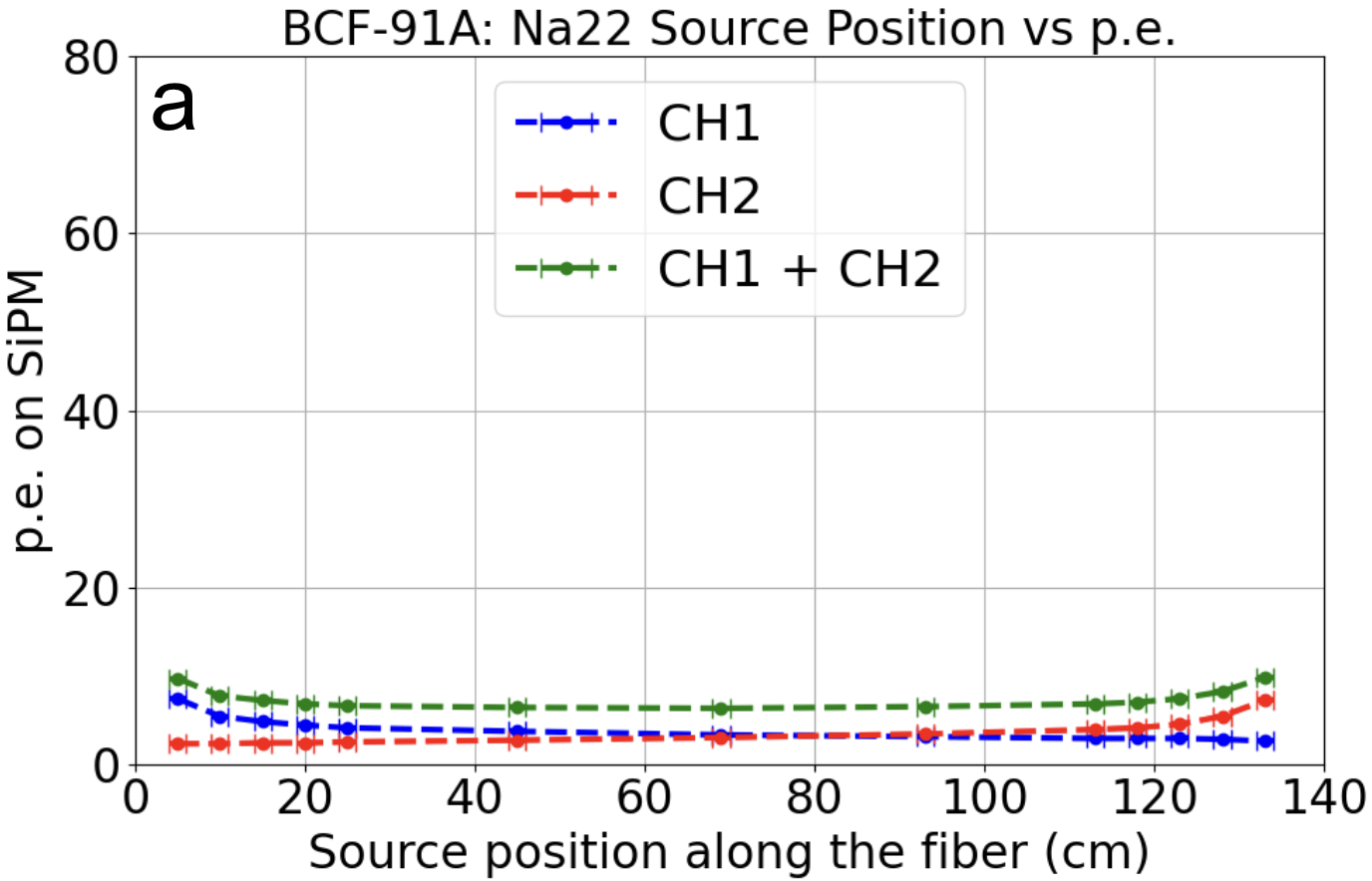}
\end{subfigure}
\hfill
\begin{subfigure}{0.495\textwidth}
  \includegraphics[width=\linewidth,height=0.65\textwidth]{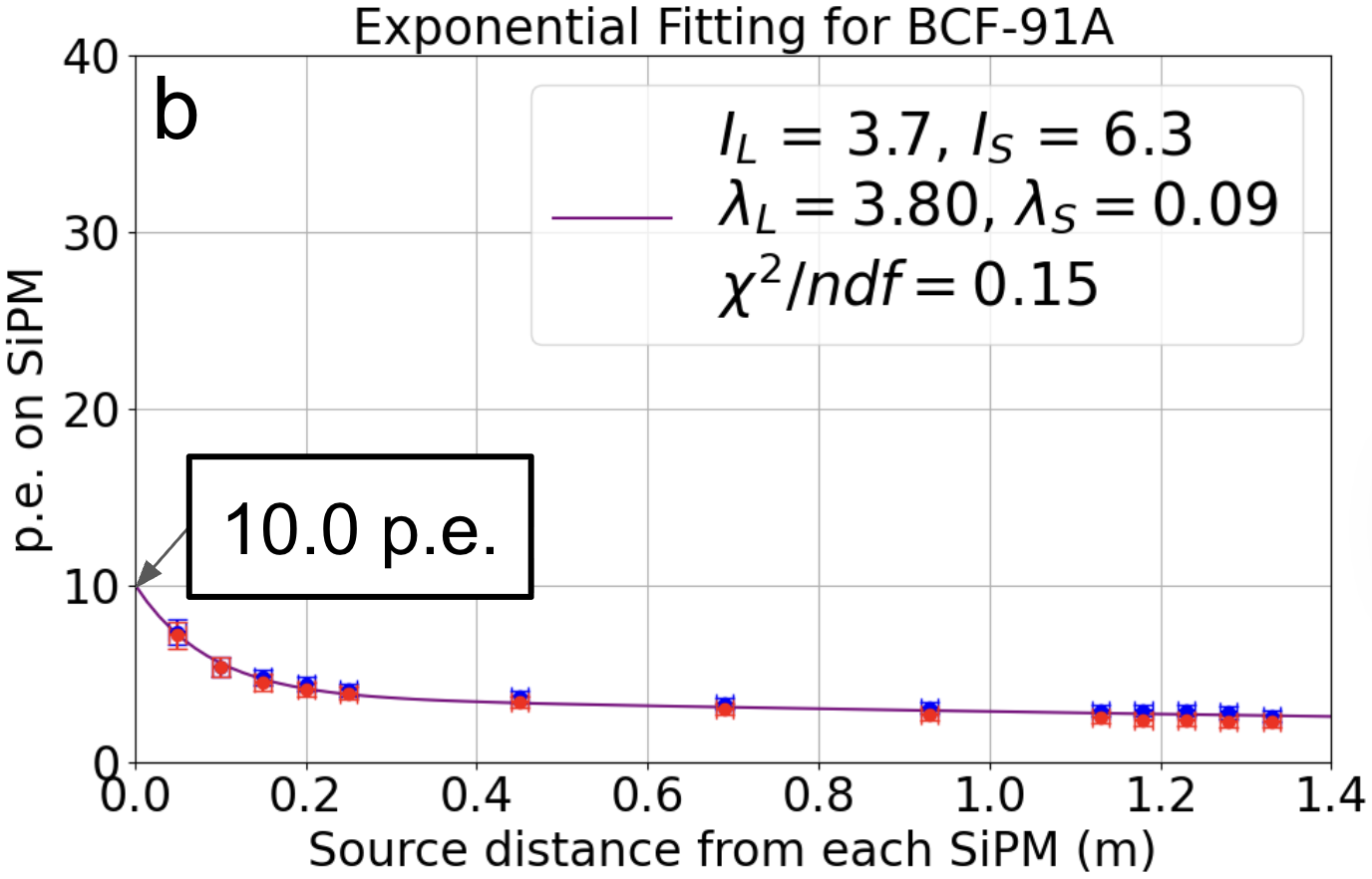}
\end{subfigure}

\vspace{1ex}

\begin{subfigure}{0.495\textwidth}
  \includegraphics[width=\linewidth,height=0.65\textwidth]{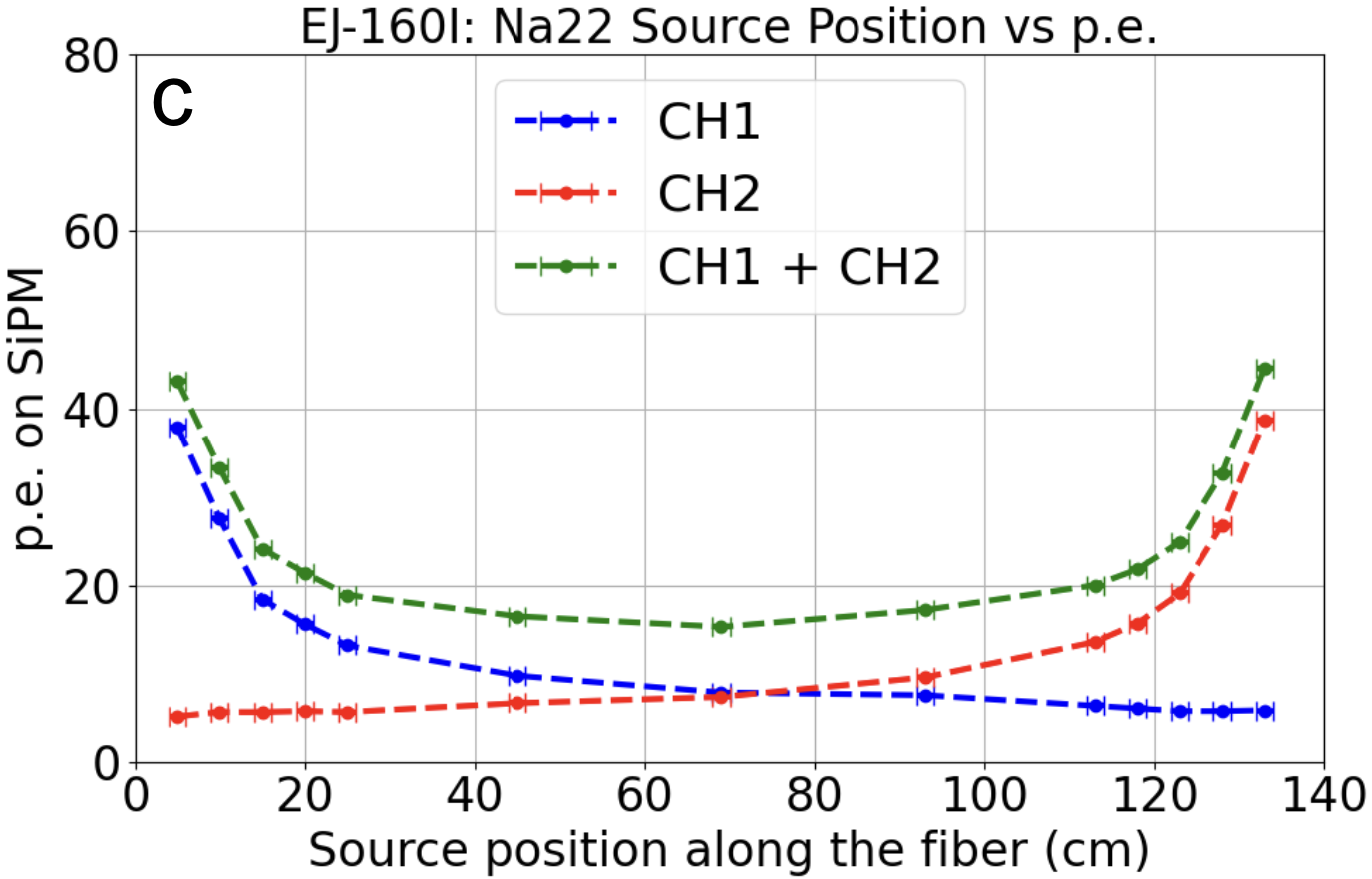}
\end{subfigure}
\hfill
\begin{subfigure}{0.495\textwidth}
  \includegraphics[width=\linewidth,height=0.65\textwidth]{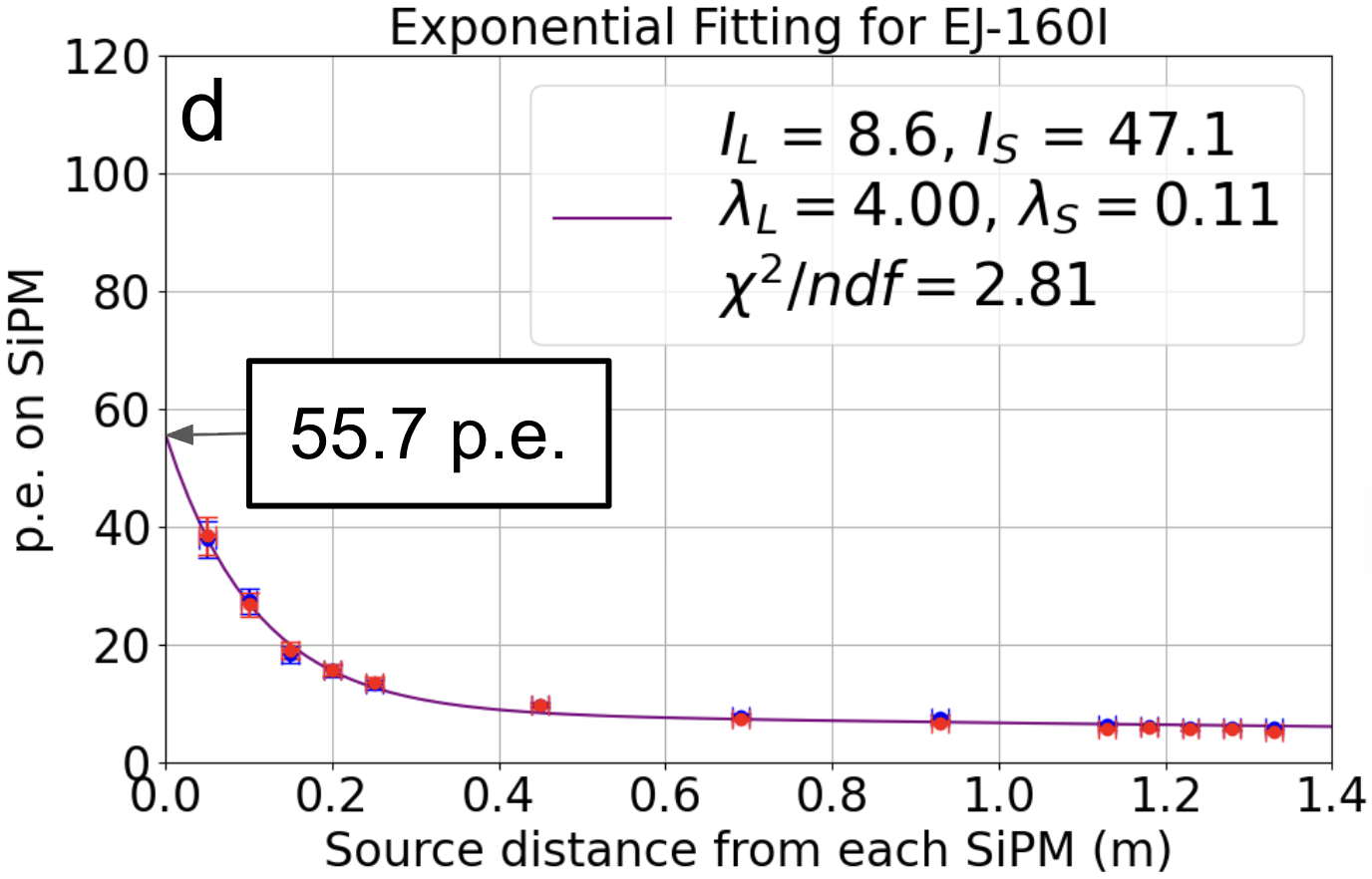}
\end{subfigure}

\vspace{1ex}

\begin{subfigure}{0.495\textwidth}
  \includegraphics[width=\linewidth,height=0.65\textwidth]{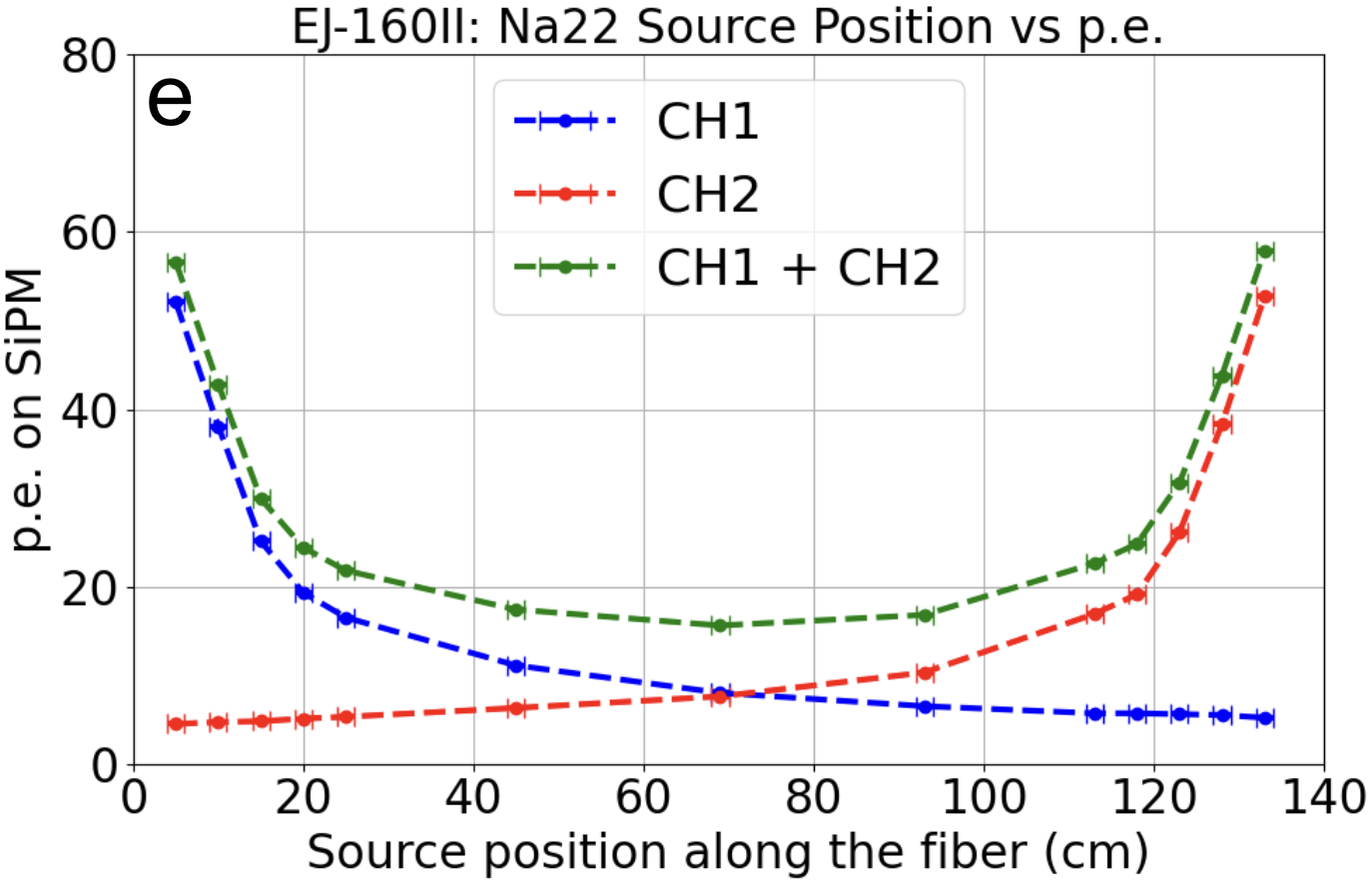}
\end{subfigure}
\hfill
\begin{subfigure}{0.495\textwidth}
  \includegraphics[width=\linewidth,height=0.65\textwidth]{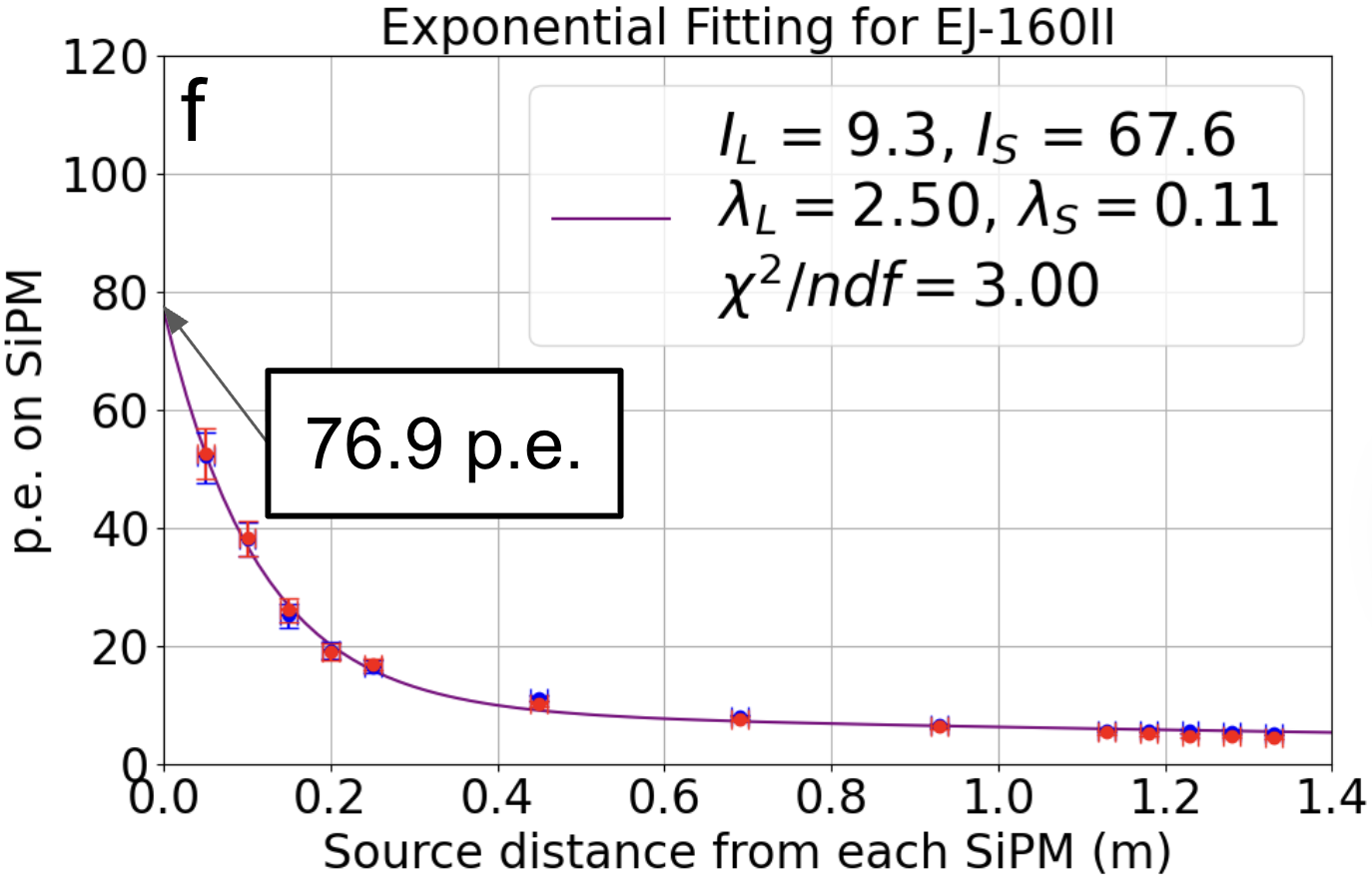}
\end{subfigure}
\caption{Light yield and fitted attenuation lengths of fibers irradiated with a \textsuperscript{22}Na $\gamma$ source. 
In the left panels, the blue, red, and green curves represent channel 1 (CH1; left SiPM), channel 2 (CH2; right SiPM), and their sum, respectively; the right panels show CH1 (blue) and CH2 (red) with the corresponding double-exponential fits.}
\label{fig: gamma_sum_fitting}
\end{figure}

\begin{figure}[h!]  
    \centering  
    \includegraphics[width=0.75\textwidth]{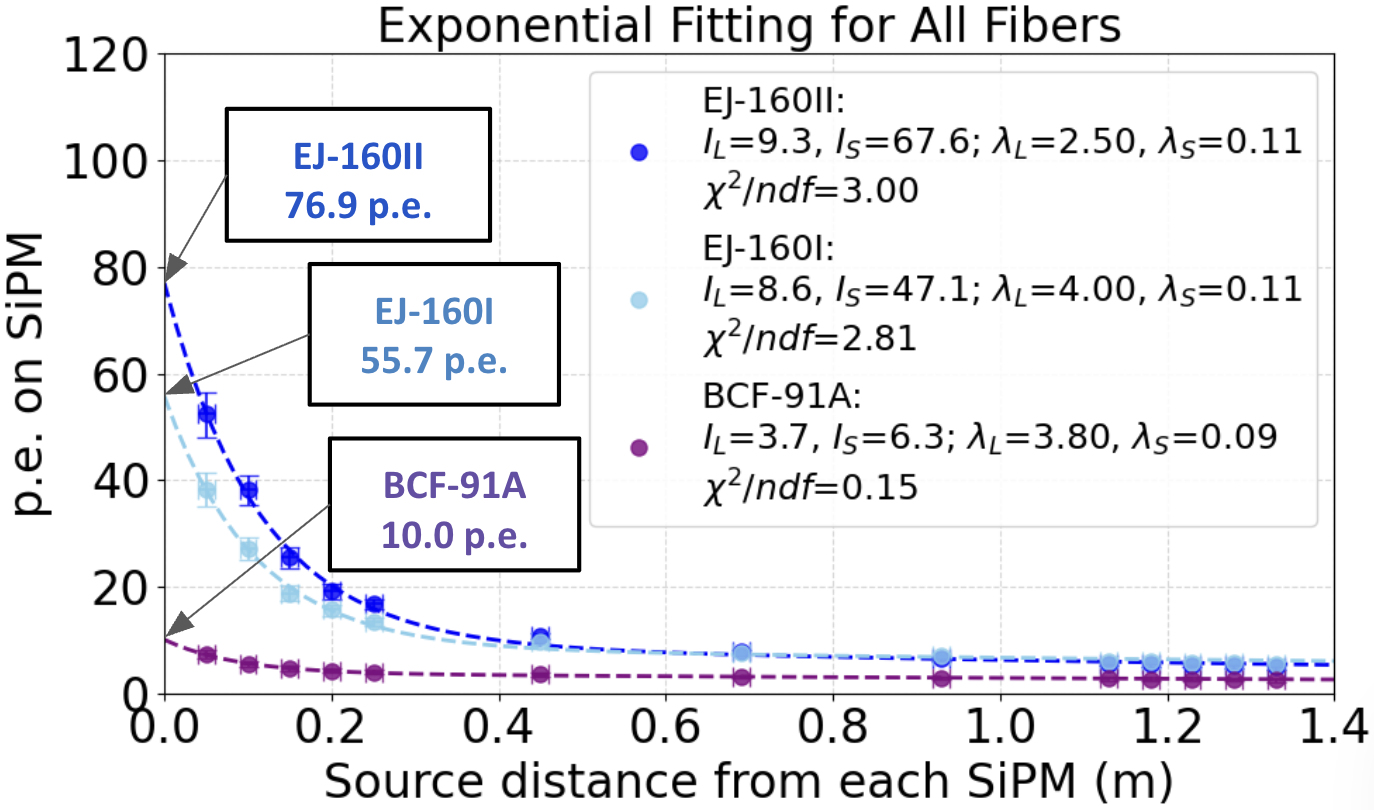}  
    \caption{\textcolor{black}{Summary of the $\gamma$ response for three tested fibers.}}  
    \label{fig: gamma_summary}  
\end{figure}

The {$\gamma$} irradiation results are shown in Figure~\ref{fig: gamma_sum_fitting}. They mirror the {$\beta$} trends. The left column of the figure, panels (a), (c) and (e), shows raw mean data, 
while the right column, panels (b), (d) and (f), shows the corresponding double-exponential fits.
The light yields at the distance of 0\,m from the SiPM, as extrapolated from the attenuation length fits, have values of 10.0\,p.e. for BCF-91A, 55.8\,p.e. for EJ-160I, and 76.9\,p.e. for EJ-160II, respectively. 
In Figure~\ref{fig: gamma_summary} we directly compare the responses to $\gamma$ of the three fibers tested.


\clearpage

\subsection{Alpha response}\label{sec:alpha_irradiation}

\begin{figure}[h!]
\centering
\includegraphics[width=.9\textwidth]{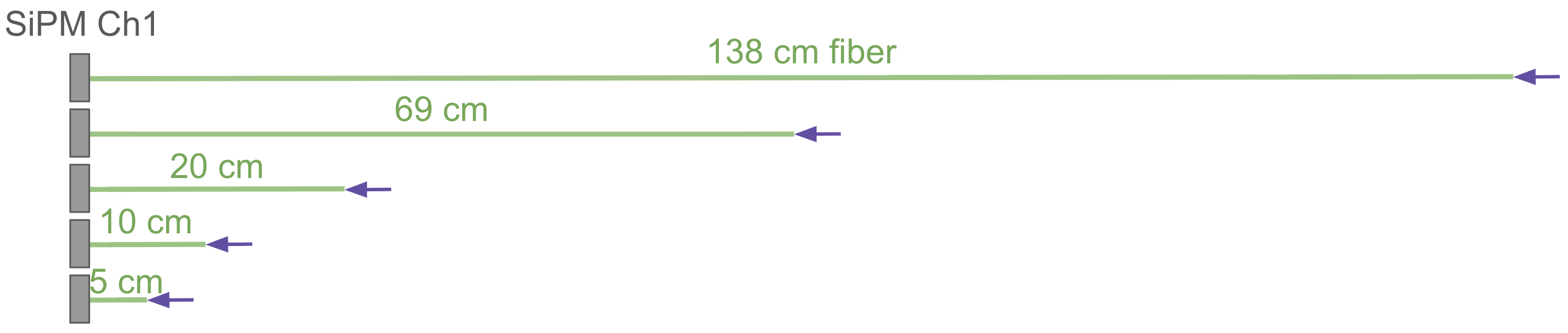}
\caption{\raggedright 
Schematic of {$\alpha$} irradiation at multiple fiber lengths with single-end SiPM coupling. 
}
\label{fig: Optical_Setup2}
\end{figure}

For these measurements, a \textsuperscript{241}Am source of activity 1.0\,$\mu$Ci was used as an {$\alpha$} point source with a predominant energy of 5.486\,MeV. 
Due to the short range of alphas (about 60--70\,$\mu$m), if irradiated transversely to the fiber most energy would be deposited in the cladding~\citep{Seltzer1993-bq-ASTAR}, so a different approach was necessary.
For each fiber type, five fibers of lengths 5, 10, 20, 69, and 138\,cm were used with both ends diamond fly-cut.
One fiber end was coupled to a single SiPM while the other end was irradiated by an {$\alpha$} source, as schematically shown in Figure~\ref{fig: Optical_Setup2}.

\begin{figure}[h!]
\centering
\captionsetup[figure]{skip=5pt}

\begin{subfigure}{0.495\textwidth}
  \centering
  \includegraphics[width=\linewidth]{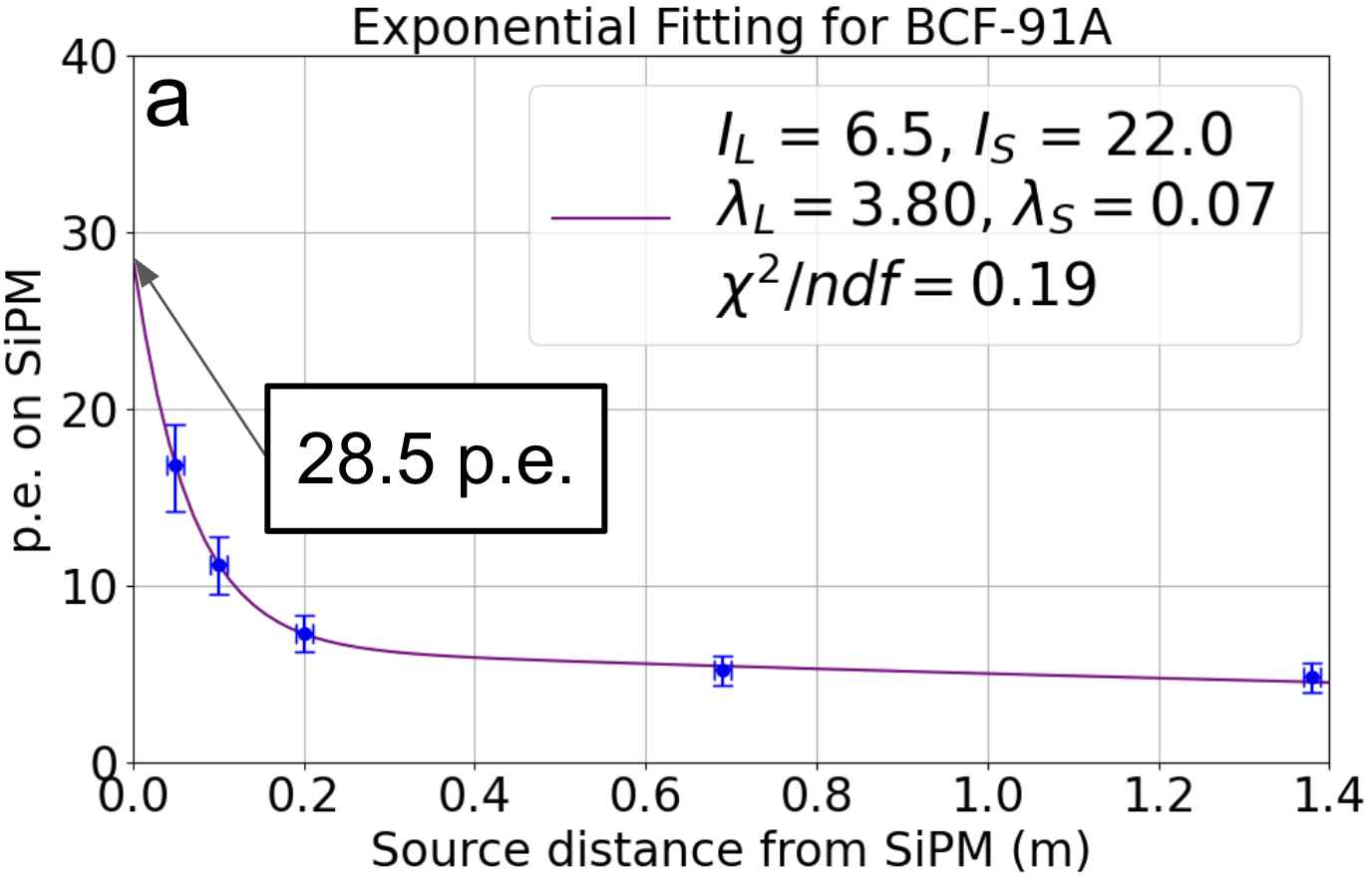}
\end{subfigure}
\hfill
\begin{subfigure}{0.495\textwidth}
  \centering
  \includegraphics[width=\linewidth]{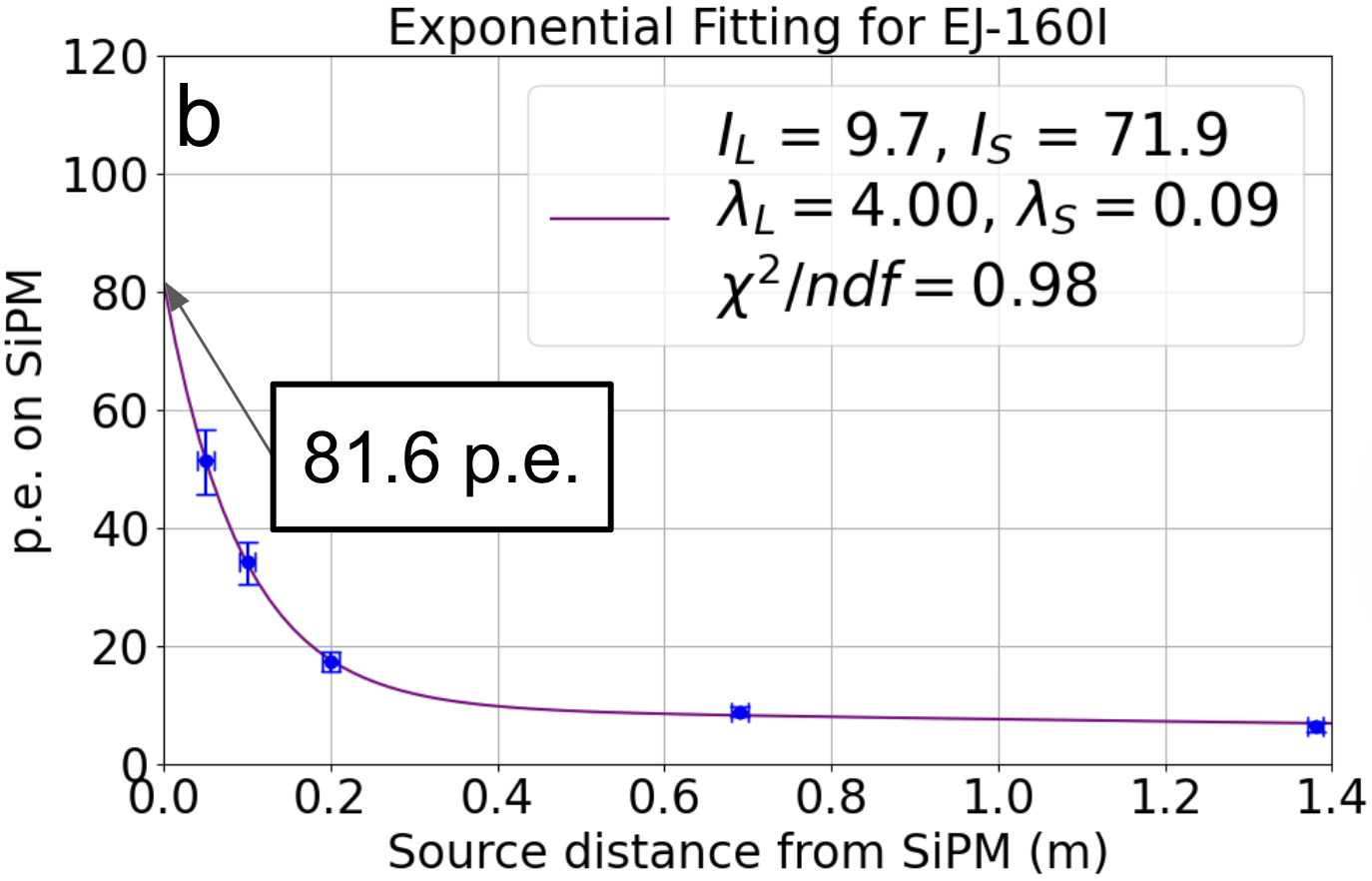}
\end{subfigure}

\vspace{1.5ex}

\begin{subfigure}{0.495\textwidth}
  \centering
  \includegraphics[width=\linewidth]{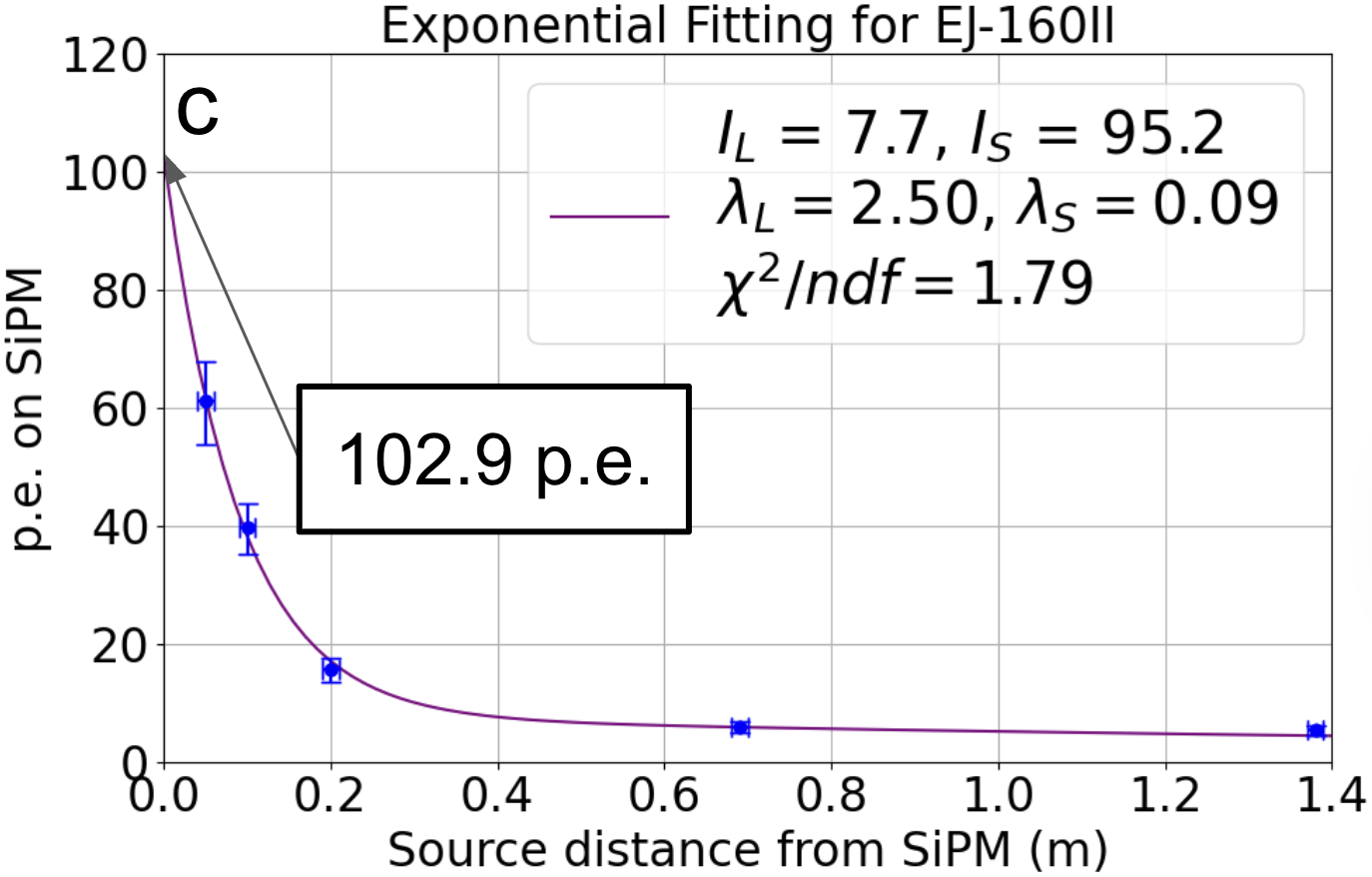}
\end{subfigure}

\caption{Light yield and fitted attenuation lengths of fibers irradiated with a \textsuperscript{241}Am $\alpha$ source.}
\label{fig: alpha fitting}
\end{figure}

\begin{figure}[h!]  
    \centering  
    \includegraphics[width=0.75\textwidth]{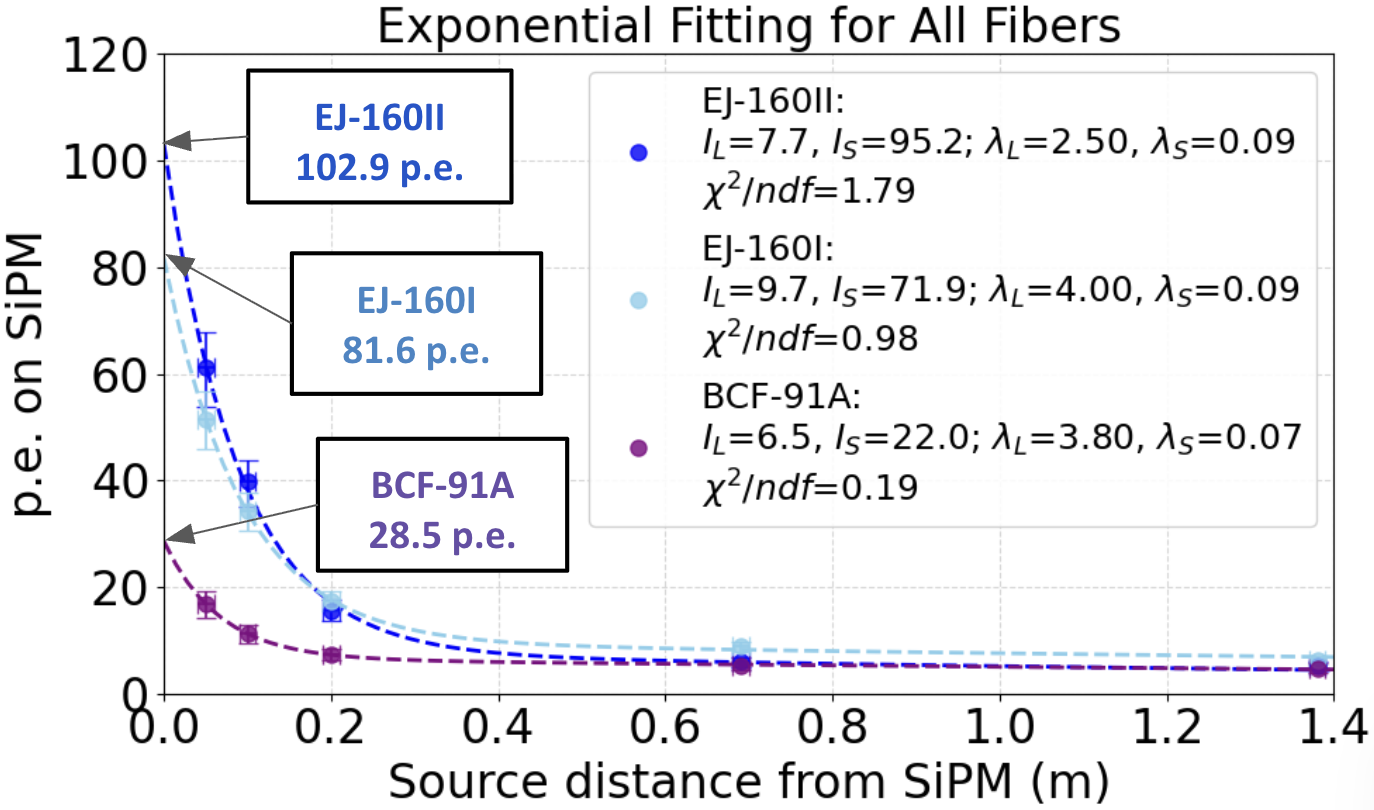}  
    \caption{\textcolor{black}{Summary of the $\alpha$ response for three tested fibers.}}  
    \label{fig: alpha_summary}  
\end{figure}

As illustrated in Figure~\ref{fig: alpha fitting}, the {$\alpha$} irradiation results show similar trends as the {$\beta$}/{$\gamma$} irradiation.
The panels (a), (b), and (c), shows raw mean data and its the corresponding double-exponential fits.
The light yields extrapolated using fitted functions to the distance 0\,m from the SiPM were
28.5\,p.e., 81.6\,p.e., and 102.9\,p.e. for BCF-91A, EJ-160I, and EJ-160II, respectively. In Figure~\ref{fig: alpha_summary} we directly compare the $\alpha$ responses of the three fibers tested.


\subsection{Summary and discussion of measurements}

Table~\ref{table: Light Detection} summarizes measurements discussed above. The table shows the number of photoelectrons extrapolated to position 0\,m using fitted double-exponential functions for each fiber and irradiation type. These light yields are quoted with systematic uncertainties that were derived from uncertainties in the placement of radioactive sources and the positioning of fibers and their couplings to SiPMs.
Statistical errors are negligible, as discussed previously.

For $\beta$ irradiation, the light yield of EJ-160I and EJ-160II is approximately 5.0 and 6.9 
times higher than that of BCF-91A, respectively. 
For $\gamma$ irradiation, the corresponding factors are 5.6 and 7.7.
For the $\alpha$ source, the enhancement is more modest, with EJ-160I and EJ-160II yielding 
factors of 2.9 and 3.6, respectively.
%
This reduction may be attributed to the stronger quenching of the scintillation light for $\alpha$ particles in the polystyrene-based fibers EJ-160I and EJ-160II~\citep{Tretyak:2010quenching}. 

\begin{table}[h!]
    \centering
    \small
    \begin{tabular}{|>{\raggedright\arraybackslash}m{2.7cm}|>{\centering\arraybackslash}m{2.5cm}|>{\centering\arraybackslash}m{3.5cm}|>{\centering\arraybackslash}m{3.5cm}|}
        \hline
        \makecell[tl]{ } & \textbf{BCF-91A} & \textbf{EJ-160I} & \textbf{EJ-160II} \\
        \hline
        \makecell[tl]{Purpose} & WLS & \makecell{Scintillation and WLS} & \makecell{Scintillation and WLS} \\
        \hline
        \makecell[tl]{Beta from \textsuperscript{90}Sr} & $(12.7 \pm 0.8)$ p.e.& $(64.0 \pm 2.9)$ p.e.& $(87.7 \pm 3.7)$ p.e.\\
        \hline
        \makecell[tl]{Gamma from \textsuperscript{22}Na} & $(10.0 \pm 0.5)$ p.e.& $(55.7 \pm 2.4)$ p.e.& $(76.9 \pm 3.2)$ p.e.\\
        \hline
        \makecell[tl]{Alpha from \textsuperscript{241}Am} & $(28.5 \pm 2.2)$ p.e.& $(81.6 \pm 3.8)$ p.e.& $(103 \pm 11)$ p.e.\\
        \hline
    \end{tabular}
    \caption{Summary of light yield of fibers irradiated with $\alpha$, $\beta$, and $\gamma$ sources.}
    \label{table: Light Detection}
\end{table}

\begin{table}[h!]
    \centering
    \small
    \begin{tabular}{|>{\raggedright\arraybackslash}m{2.7cm}|>{\centering\arraybackslash}m{3.5cm}|>{\centering\arraybackslash}m{3.5cm}|>{\centering\arraybackslash}m{3.5cm}|}
        \hline
        \makecell[tl]{ } & \textbf{BCF-91A} & \textbf{EJ-160I} & \textbf{EJ-160II} \\
        \hline
        \makecell[tl]{Beta from \textsuperscript{90}Sr} 
        & \makecell{$\lambda_{\text{long}}$ 
        = 3.80\,m\\ 
        $\lambda_{\text{short}}$ = $(0.10 \pm 0.02)$\,m}
        & \makecell{$\lambda_{\text{long}}$ 
        = 4.00\,m\\ 
        $\lambda_{\text{short}}$ = $(0.11 \pm 0.01)$\,m}
        & \makecell{$\lambda_{\text{long}}$ 
        = 2.50\,m\\ 
        $\lambda_{\text{short}}$ = $(0.12 \pm 0.01)$\,m}
        \\
        \hline
        
        \makecell[tl]{Gamma from \textsuperscript{22}Na} 
        & \makecell{$\lambda_{\text{long}}$ 
        = 3.80\,m\\ 
        $\lambda_{\text{short}}$ = $(0.09 \pm 0.01)$\,m} 
        & \makecell{$\lambda_{\text{long}}$ 
        = 4.00\,m\\ 
        $\lambda_{\text{short}}$ = $(0.11 \pm 0.01)$\,m} 
        & \makecell{$\lambda_{\text{long}}$ 
        = 2.50\,m\\ 
        $\lambda_{\text{short}}$ = $(0.11 \pm 0.01)$\,m}
        \\
        \hline
        
        \makecell[tl]{Alpha from \textsuperscript{241}Am} 
        & \makecell{$\lambda_{\text{long}}$ 
        = 3.80\,m\\ 
        $\lambda_{\text{short}}$ = 0.07\,m} 
        & \makecell{$\lambda_{\text{long}}$ 
        = 4.00\,m\\ 
        $\lambda_{\text{short}}$ = 0.09\,m} 
        & \makecell{$\lambda_{\text{long}}$ 
        = 2.50\,m\\ 
        $\lambda_{\text{short}}$ = 0.09\,m} \\
        \hline
    \end{tabular}
    \caption{Summary of attenuation lengths of fibers irradiated with $\alpha$, $\beta$, and $\gamma$ sources.
    As discussed in the text, the values of $\lambda_{\text{long}}$ are taken from~\cite{Bae:2025fiberLED} due 
    to the short lengths of tested fibers. 
    }
    \label{table: Attenuation length}
\end{table}

Table~\ref{table: Attenuation length} summarizes the attenuation lengths obtained for each fiber and irradiation type.
In the double-exponential fits, all measured profiles exhibit a pronounced short component. This behavior is well known in scintillating and WLS fibers; it is attributed to distinct core-guided and cladding-guided light propagation.
While the cladding-guided component has a significantly larger initial contribution—typically by a factor of 3–4 due to the larger trapping efficiency of the cladding—it attenuates rapidly within the first meter~\cite{Amos1990:fiber-cladding_light, Achenbach:fiber-cladding_light}.
Consequently, $\lambda_{\text{short}}$ mainly reflects this cladding-guided contribution, whereas $\lambda_{\text{long}}$ is associated with the core-guided component.

Accordingly, it is often practical to report $\lambda_{\text{long}}$ over restricted ranges that exclude the short-distance region (e.g., 1.0--2.8\,m~\cite{Barbosa2013} and 1.0--4.0\,m~\cite{Zhang:2019LHAASO-AL}), and manufacturers (e.g., Kuraray~\cite{Kuraray}) commonly quote a single attenuation length consistent with $\lambda_{\text{long}}$.
In the present work, the fibers are approximately 1.4\,m long, matching the LEGEND-1000 design. This limited length provides insufficient leverage to constrain $\lambda_{\text{long}}$ independently. 
Therefore, we adopt the $\lambda_{\text{long}}$ values determined from a related study using the same fiber types but with lengths of nearly 3.0\,m~\cite{Bae:2025fiberLED} and fit only $\lambda_{\text{short}}$ to model the measured position dependence.

We observe that although WLS fibers are not designed for scintillation, they produce detectable signals when irradiated by ionizing radiation. This is consistent with intrinsic scintillation in aromatic polymers such as polystyrene and polyvinyltoluene~\cite{Chakraborty:2017, Nakamura:2015}. 

\section{Conclusions}

We have characterized the radiation response of two new scintillating-wavelength-shifting fibers, EJ-160I and EJ-160II, under $\alpha$, $\beta$ and $\gamma$ irradiation, and compared their performance to that of the wavelength-shifting fiber BCF-91A. 
The EJ-160I and EJ-160II fibers yield approximately five and seven times more photoelectrons at the SiPM readout, respectively, compared to BCF-91A. 
The fitted attenuation lengths are 3.80\,m for BCF-91A, 4.00\,m for EJ-160I and 2.50\,m for EJ-160II. When comparing the EJ-160 variants, EJ-160II shows higher light yield with a shorter attenuation length compared to the EJ-160I.

This publication presents partial results of the ongoing program to develop better Sci-WLS fibers and improve their radio purity.
We also advance a comprehensive simulation framework to model light yield and photon transport in WLS and Sci-WLS fibers that will further benefit this work.
A similar approach was previously validated for Kuraray Y-11 fibers~\cite{Pahlka2019}.
As part of our broader ongoing program on fiber characterization, we plan to examine additional commercially available WLS fibers (e.g., Kuraray~\cite{Kuraray}). 
These studies, together with continued fiber development and modeling, will be reported separately.


\acknowledgments{
We thank Prof.~S.~Schönert, Dr.~P.~Krause, and the group at the Technical University of Munich for providing BCF-91A fiber samples and for insightful discussions. 
This work was supported in part by the University of Texas at Austin and the U.S. National Science Foundation under grant PHY-2312278.
}


\providecommand{\href}[2]{#2}\begingroup\raggedright\endgroup

\end{document}